\documentclass[11pt,titlepage]{article}
\usepackage[margin=1.2in]{geometry}
\usepackage{amsmath, amsbsy, amsthm, array, mathrsfs}
\usepackage{graphics}
\usepackage{physics}
\usepackage{epsfig, amssymb,latexsym,verbatim}
\usepackage{graphicx}
\usepackage{color}
\usepackage{dsfont, bbm}
\usepackage{relsize}
\usepackage{subcaption}

\newcommand{\R}{\mathbb{R}}

\newcommand{\noin}{\noindent}
\newcommand{\bee}{\begin{eqnarray*}}
\newcommand{\ene}{\end{eqnarray*}}
\newcommand{\bec}{\begin{center}}
\newcommand{\enc}{\end{center}}
\newcommand{\be}{\begin{equation}}
\newcommand{\ee}{\end{equation}}

\newcommand{\mc}{\mathcal}

\newcommand{\ep}{\varepsilon}
\newcommand{\mb}{\mathbf}
\newcommand{\bs}{\boldsymbol}
\newcommand{\tb}{\textbf}
\newcommand{\pend}{$\blacksquare$}
\newcommand{\vs}{\vskip 3mm}
\newcommand{\bi}{\begin{itemize}}
\newcommand{\ei}{\end{itemize}}

\begin{document}
\baselineskip 3.2ex

\title{\LARGE Least sum of squares of trimmed residuals regression} 
\vs
\vs
\author{ {\sc Hanwen Zuo and
Yijun Zuo}\\[2ex]
         {\small {\em  Department of Computer Science} and
         {\em Department of Statistics and Probability} }\\[.5ex]
         {\small Michigan State University, East Lansing, MI 48824, USA} \\[2ex]
         {\small zuohanwe@msu.edu and
         zuo@msu.edu}\\[6ex]
     }
 \date{\today}
\maketitle

\vskip 3mm
{\small

\begin{abstract}
In the famous least sum of trimmed squares (LTS) of residuals estimator (Rousseeuw (1984)), residuals are first squared and then trimmed. In this article, we first trim residuals - using a depth trimming scheme - and then square the rest of residuals. The estimator that can minimize the sum of squares of the trimmed residuals, is called an LST estimator.
\vs
It turns out that LST is a robust alternative to the classic least sum of squares (LS)  estimator.  Indeed, it has a very high finite sample breakdown point, and can resist, asymptotically, up to $50\%$ contamination without breakdown - in sharp contrast to the $0\%$ of the LS estimator.
\vs
The population version of LST is Fisher consistent, and the sample version is strong and root-$n$ consistent and asymptotically normal.
 Approximate algorithms for computing LST are proposed and tested in synthetic and real data examples. These experiments indicate that one of the algorithms can compute LST estimator very fast and with relatively smaller variances, compared with that of the famous LTS estimator. All the evidence suggests that LST deserves to be a robust alternative to the LS estimator and is feasible in practice for high dimensional data sets (with possible contamination and outliers).




\bigskip
\par

\bigskip
\par
\noindent {\bf Running title:} the least squares of trimmed residuals.
\end{abstract}
}
\setcounter{page}{1}

\section {Introduction}
\begin{figure}[h]
    \centering
    \begin{subfigure}[t]{0.47\textwidth}
    \includegraphics[width=\textwidth]{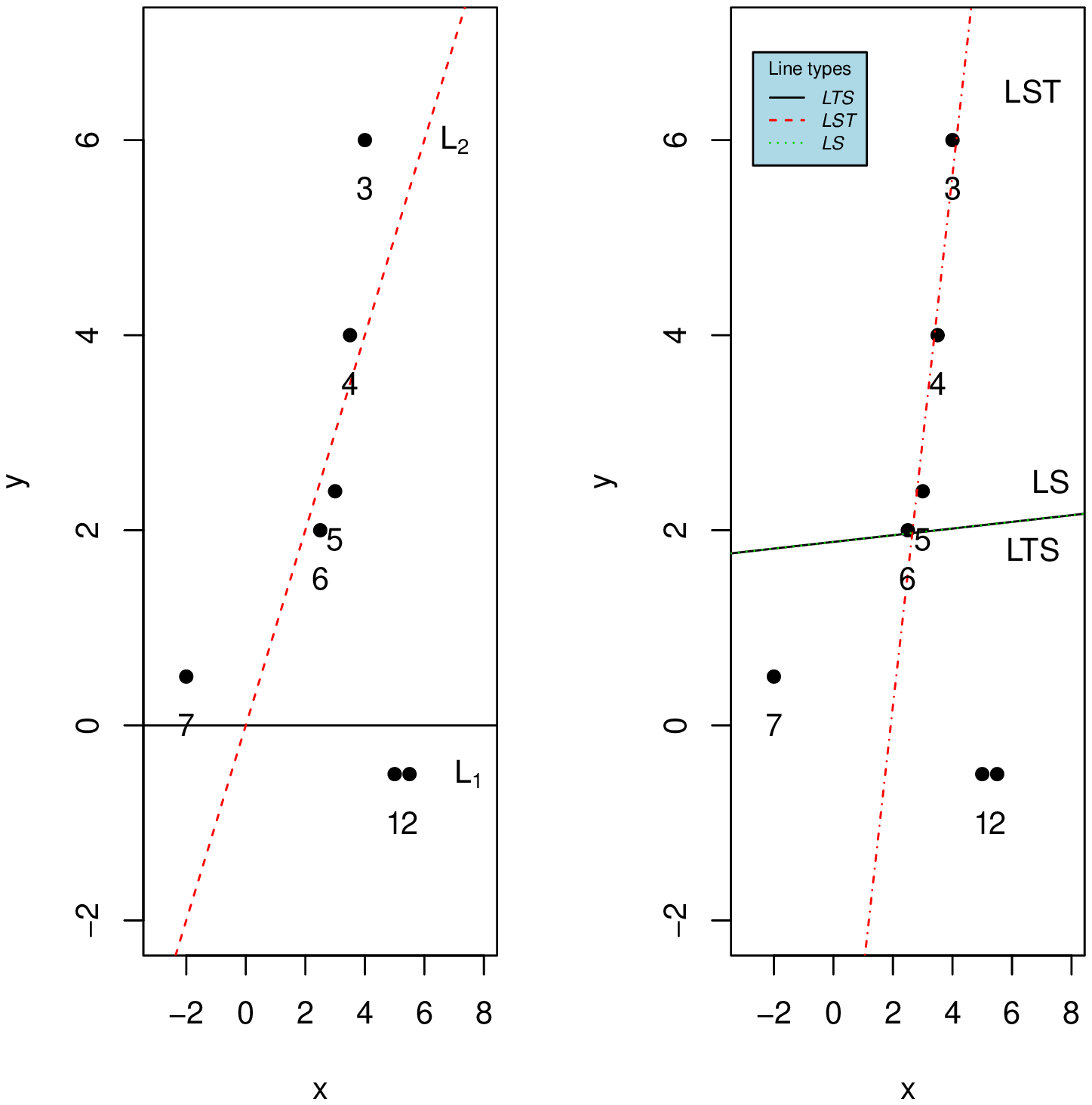}            
        \caption{Left panel: plot of seven artificial points and two candidate lines ($L_1$ and $L_2$), which line would you  pick?
        Sheerly based on the trimming scheme and objective function value, if one uses the number $h=\lfloor n/2\rfloor+\lfloor (p+1)/2\rfloor$ given on page 132 of RL87, that is, employing four squared residuals, then LTS prefers $L_1$ to $L_2$ whereas LST reverses the preference.\\[1ex] Right panel:  the same seven points are fitted by LTS,  LST, and the LS (benchmark). A solid black  line is LTS given by ltsReg. Red dashed line is given by LST, and green dotted line is given by the LS - which is identical to LTS line in this case.}
        \label{fig:seven-points-lines}
     \end{subfigure}
     \hspace*{2mm}
    \begin{subfigure}[ht]{0.47\textwidth}
    \vspace*{-19mm}
    \includegraphics[width=\textwidth]{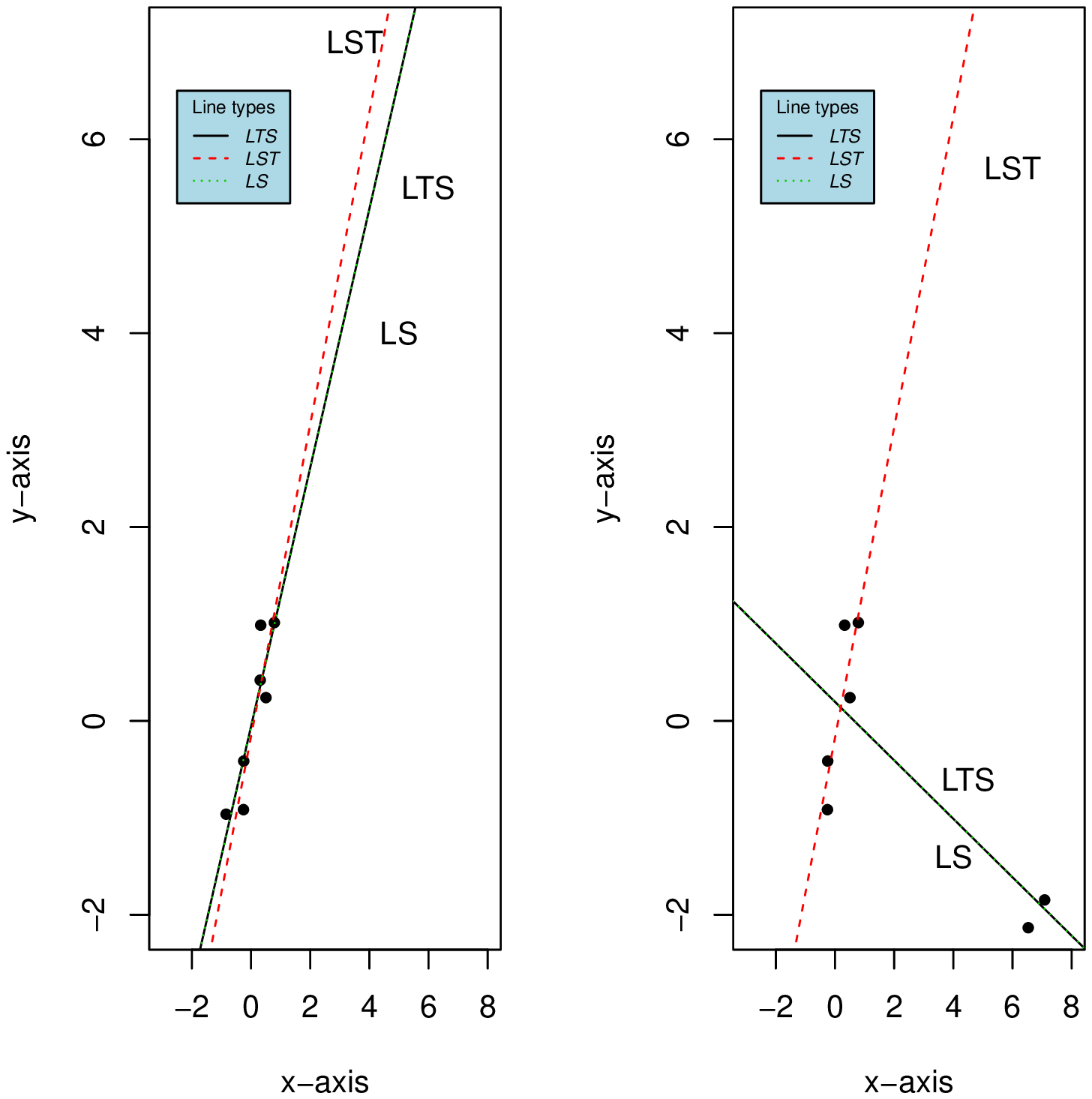}  
        \caption{Left panel: plot of seven highly correlated normal points (with mean being the zero vector and covariance matrix with diagonal entries being one and off-diagonal entries being 0.88) and three lines given by LST , LTS, and LS. LS line is identical to LTS line again.\\[1ex] 
        Right panel: LTS line (solid black) and LST line (dashed red), and LS (dotted green) for the same seven highly correlated normal points but with two points contaminated nevertheless. The LS line is identical to LTS line due to the attributes in the R function ltsReg that is based on Rousseeuw and Van Driessen (2006) (RVD06)).}
        \label{fig:lts-lines}
   \end{subfigure}
   \caption{(a) Difference between the two procedures: LST and LTS. (b) Performance difference between LST and LTS when there are contaminated points ($x$-axis leverage points).}
   \label{Fig:one}
\vspace*{-10mm}
\end{figure}

In the classical regression analysis, we assume that there is a relationship for a given data set $\{(\bs{x}'_i, y_i)', i=1,\cdots, n\}$:
\be
y_i=(1,\bs{x}'_i)\bs{\beta}_0+{e}_i,~~ i=1,\cdots, n,  \label{model.eqn}
\ee
where $y_i\in \R^1$, $'$ stands for the transpose, $\bs{\beta}_0=(\beta_{01}, \cdots, \beta_{0p})'$ (the true unknown parameter) in $\R^p$ and~ $\bs{x_i}=(x_{i1},\cdots, x_{i(p-1)})'$ in $\R^{p-1}$, $e_i\in \R^1$ is called an error term (or random fluctuation/disturbances). That is, $\beta_{01}$ is the intercept term of the model. Write $\bs{w}_i=(1,\bs{x}'_i)'$, then one has $y_i=\bs{w}'_i\bs{\beta}_0+e_i$, which will be used interchangeably with model (\ref{model.eqn}).
\vs
We like to estimate the $\bs{\beta}_0$ based on a given sample $\mb{Z}^{(n)}
:=\{(\bs{x}'_i, y_i)', i=1,\cdots, n\}$ from the model $y=(1,\bs{x}')\bs{\beta}_0+e$.
 Call the difference between $y_i$ and $\bs{w'_i}{\bs{\beta}}$
 the ith residual, $r_i$, for a candidate coefficient vector $\bs{\beta}$ (which is suppressed). That is,
\be {r}_i=y_i-\bs{w'_i}{\bs{\beta}}.\label{residual.eqn}
\ee
To estimate $\bs{\beta}_0$, the classic \emph{least squares} (LS) estimator is the minimizer of the sum of the squared residuals
$$\widehat{\bs{\beta}}_{ls}=\arg\min_{\bs{\beta}\in\R^p} \sum_{i=1}^n r^2_i. $$
Alternatively, one can replace the square above by absolute value to obtain the least absolute deviations  estimator (aka, $L_1$ estimator, in contrast to the $L_2$ (LS) estimator).
\vs
The least-squares estimator is  most popular  in practice across a broader spectrum of disciplines due to its great computability and optimal properties when the error $e_i$ follows a normal $\mc{N}(\mu,\sigma^2)$ distribution.
It, however, can behave badly when the error distribution is slightly departed from the normal distribution,
particularly when the errors are heavy-tailed or contain outliers.
\vs
Robust alternatives to the $\widehat{\bs{\beta}}_{ls}$ abound in the literature for a long time. The most popular ones are, among others,  M-estimators (Huber(1964)), least median squares (LMS) and least trimmed squares (LTS) estimators (Rousseeuw (1984)),  S-estimators (Rousseeuw and Yohai (1984)), MM-estimators (Yohai (1987)), $\tau$-estimators (Yohai and Zamar (1988)) and maximum depth estimators (Rousseeuw and Hubert (1999) and Zuo (2021a, 2021b)). For more related discussions, please see, Sections 1.2 and 4.4 of Rousseeuw and Leroy (1987) (RL87), and  Section 5.14 of Maronna, Martin, and Yohai (2006) (MMY06).
\vs
Among all robust alternatives,  in practice, LTS is one of the most prevailing crossing multiple disciplines. Its idea is simple, ordering the squared residuals and then trimming the larger ones and keeping at least $\lfloor n/2\rfloor$ squared residuals, where $\lfloor ~\rfloor$ is the floor function, the  minimizer of the sum of those \emph{trimmed squared residuals} is called the LTS estimator:
\[
\widehat{\bs{\beta}}_{lts}:=\arg\min_{\bs{\beta}\in \R^p} \sum_{i=1}^h (r^2)_{i:n},
\]
where $(r^2)_{1:n}\leq (r^2)_{2:n}\leq \cdots, (r^2)_{n:n}$ are the ordered squared residuals and $\lfloor n/2\rfloor \leq h \leq n$.
\vs
One naturally wonders, what if one first trims (employing the scheme given in Section 2) the residuals and then minimizes the sum of \emph{squares of trimmed residuals} (the minimizer will be called LST)? Is there any difference between the two procedures?
Outlying (extremely large or small) original residuals are trimmed after squaring in LTS - those residuals certainly are trimmed in LST. But the outlying residuals which have a small squared magnitude will not be trimmed in LTS and are trimmed in LST (see (a) of Figure \ref{Fig:one}).  Before formally introducing LST in Section 2, let us first appreciate the difference between the two procedures.
\vspace*{-5mm}
\bec
\vspace*{-0mm}
\enc
\vs
\noindent
\tb{Example 1.1} ~~We constructed a small data set in $\R^2$ with $x=(5, 5.5, 4, 3.5, 3, 2.5, -2)$ and $y=(-.5, -.5, 6, 4, 2.4, 2, .5)$, they are plotted in the left panel of the (a)
 of Figure \ref{Fig:one} above. We also provide two candidate regression lines $L_1$ ($y=0$) and $L_2$ ($y=x$). Which one would you pick to represent the overall pattern of the data set?\vs
 If one uses the number $h=\lfloor n/2\rfloor+\lfloor (p+1)/2\rfloor$ given on page 132 of RL87 to  achieve the maximum possible breakdown point (see Section 3 for definition) for LTS estimator, that is, employing four smallest squared residuals, then LTS prefers $L_1$ (using residuals from points $1$, $2$, $6$, and $7$) to $L_2$ (using points $4,5,6,7$), whereas for LST,  $L_2$ (using residuals from points $4,5,6,7$) 
  is the preferred. One might immediately argue that this is not representative since LTS searches all possible (not just two) lines and outputs the best one.

 \vs
If one utilized the R function ltsReg, then it produced the solid (black) line  whereas the line based on algorithms (see Section 5) for LST is the dashed (red) one in the right panel of the (a) of Figure \ref{Fig:one}. For benchmark purposes, the LS line  dotted (green) is also given, which is overlapping with LTS line.
From this instance, One can appreciate the difference between trimming schemes of LTS and LST. Of course, one might argue that the data set in the (a) is purely synthetic.
\vs
So, in the (b) of Figure \ref{Fig:one}, we generated seven highly correlated normal points (with correlation $0.88$ between $x$ and $y$), when there is no contamination LTS (identical to the LS again) and LST pick  perfectly  the linear pattern
whereas if there are two contaminated points (note that LTS allows $m:=\lfloor(n-p)/2\rfloor=2$ contaminated points in this case in light of Theorem 6 on page 132 of RL87), the line from LTS drastically changes in this particular instance, which again is identical to the LS. 
\vs
For examples with an increased sample size, see Section 6. Incidentally, the instability of LMS (not the LTS) was already documented in Hettmansperger and Sheather (1992). \hfill \pend
\vs
The rest of the article is organized as follows. Section 2 introduces trimming schemes and the least sum of squares of trimmed (LST) residuals estimator and establishes the existence and equivariance properties.  Section 3 investigates the robustness of LST in terms of its finite sample breakdown point and its influence function. Section 4 establishes the Fisher as well as the strong and the root-n consistency. The asymptotic normality is derived from stochastic equicontinuity  in Section 5. Section 6 is devoted to the computation algorithms of LST where two approximate algorithms are proposed. Section 7 presents examples of simulated and real data and carries out the comparison with the leading regression estimator, LTS. Section 8 consists of some concluding discussions. 
Long proofs are deferred to the Appendix.

\section{Least sum of squares of trimmed residuals estimator}
\subsection{Trimming schemes}

\tb{Rank based trimming} ~~
This scheme is based on the ranks of data points, usually trimming an equal number of points at both tails of a data set (that is, lower or higher rank points are trimmed) and also can trim points one-sided if needed (such as when all data points lie on the positive (or negative) side of number axis).
\vs
This scheme is closed related to the trimmed mean,  which can keep a good balance between robustness and efficiency,
alleviating the extreme sensitivity of sample mean and  enhancing the efficiency of the sample median. Trimmed mean
 has been used in practice for more than two centuries (see Hampel, Ronchetti, Rousseeuw and Stahel (1986) (HRRS86) 1986, page 34), and is attributed to ``Anonymous" (1821) (Gergonne, see Stigler, 1976), or Mendeleev, 1895. Tukey (Tukey and McLaughlin (1963), Dixon and Tukey (1968)) is one of the outstanding advocators for the trimmed mean in the last century.
\vs
Rank-based trimming focuses only on the relative position of points with respect to others and ignores the magnitude of the point and the relative distance between points.
Zuo (2006) and Wu and Zuo (2009) discussed an alternative trimming scheme, which
 exactly catches these two important attributes (magnitude and relative distance). It orders data from a center (the median) outward and trims the points that are far away from the center. This is known as depth-based
trimming.
\vs
\noin
\tb{Depth (or outlyingness) based trimming}~~
In other words, the depth-based trimming
scheme trims points that lie on the outskirts (i.e. points that are less deep, or outlying). The depth (or,
equivalently, outlyingness ) of a point x is defined to be
\be
D(x, X^{(n)})=|x-\mbox{Med}(X^{(n)})|/\mbox{MAD}(X^{(n)}),  \label{outlyingness.eqn}
\ee
where $X^{(n)}=\{x_1, \cdots, x_n\}$ is a data set in $\R^1$,  Med$(X^{(n)})=\mbox{median}(X^{(n)})$ is the median of the data points, and  MAD$(X^{(n)})=\mbox{Med}(\{|x_i-\mbox{Med}(X^{(n)})|,~ i=1,2, \cdots, n\})$ is the median of absolute deviations to the center (median). It is readily seen that $D(x, X^{(n)})$ is a generalized standard deviation, or equivalent to the one-dimensional projection depth/outlyingness (see Zuo and Serfling (2000) and Zuo (2003,2006) for a high dimensional version). For notion of outlyingness, cf  Stahel (1981), Donoho (1982), and Donoho and Gasko (1992).
\vs
LTS essentially employs one-sided rank based trimming scheme (w.r.t. squared residuals), whereas depth based trimming is utilized in LST which is introduced next.\vs

\subsection{Definition and properties of LST}
\vs\noindent
\tb{Definition}~
For a given sample $\mb{Z}^{(n)}=\{(\bs{x}'_i, y_i)', 1\leq i\leq n\}$ in $\R^{p}$ from $y=\bs{w}'\bs{\beta}_0+e$ and a $\bs{\beta} \in \R^p$, define
\begin{align}
m_n(\bs{\beta}):=m(\bs{Z}^{(n)},\bs{\beta})&=\mbox{Med}_i\{r_i\}, \label{med.eqn}\\
\sigma_n(\bs{\beta}):=\sigma(\bs{Z}^{(n)},\bs{\beta})&=\mbox{MAD}_i\{r_i\}, \label{mad.eqn}
\vspace*{-8mm}
\end{align}

\noin
where operators Med and MAD are used for discrete data sets (and distributions as well) and $r_i$ defined in (\ref{residual.eqn}).   
For a constant $\alpha$ in the depth trimming scheme, consider the quantity
\be
Q(\bs{Z}^{(n)}, \bs{\beta}, \alpha):=\sum_{i=1}^{n}r_i^2\mathds{1}\left( \frac{|r_i-m(\bs{Z}^{(n)},\bs{\beta})|}{\sigma(\bs{Z}^{(n)},\bs{\beta})}\leq \alpha\right),\label{objective.eqn}
\ee
 where $\mathds{1}(A)$ is the indicator of $A$ (i.e., it is one if A holds and zero otherwise).
Namely, residuals with their depth (or outlyingness) greater than $\alpha$ will be trimmed.
When there is a majority ($\geq \lfloor(n+1)/2\rfloor$) identical $r_i$s, we define $\sigma(\mb{Z}^{(n)}, \bs{\beta})=1$ (since those $r_i$  lie in the deepest position (or are the least outlying points)).
\vs
Minimizing $Q(\bs{Z}^{(n)}, \bs{\beta}, \alpha)$, one gets the \emph{least} sum of \emph{squares} of {\it trimmed} (LST) residuals estimator,
\be
\widehat{\bs{\beta}}^n_{lst} :=\widehat{\bs{\beta}}_{lst}(\mb{Z}^{(n)}, \alpha)=\arg\min_{\bs{\beta}\in \R^p}Q(\bs{Z}^{(n)}, \bs{\beta}, \alpha).\label{lst.eqn}
\ee
One might take it for granted that the  minimizer of  $Q(\bs{Z}^{(n)}, \bs{\beta}, \alpha)$ always exists.
Does the right-hand side (RHS) of (\ref{lst.eqn}) always have a minimizer? If it exists, is it unique? We treat this problem formally next.
Assume $\mb{X}_n=(\bs{w}_1, \cdots, \bs{w}_n)'$ has a full rank $p$ ($p<n$) throughout.
\vs

Hereafter we will assume that $\alpha\geq 1$. That is, we will keep the residuals that are no greater than one MAD away from the center (the median of residuals) untrimmed.  For a given $\alpha$, $\bs{\beta}$, and $\bs{Z}^{(n)}$, define a set of indexes for $1\leq i\leq n$
\be
I(\bs{\beta})=\Big\{ i: \frac{|r_i-m(\bs{Z}^{(n)},\bs{\beta})|}{\sigma(\bs{Z}^{(n)},\bs{\beta})}\leq \alpha\Big\}. \label{I-beta.eqn}
\ee
Namely, the set of subscripts so that the depth (see (\ref{outlyingness.eqn})) of the corresponding residuals are no greater than $\alpha$. It depends on $\mb{Z}^{(n)}$ 
 and
 $\alpha$, which are suppressed  in the notation. 
Following the convention, 
we denote the cardinality of set $A$ by $|A|$. We have
\vs
\noin
\tb{Lemma 2.1} For any $\bs{\beta}\in \R^p$ and the given $\mb{Z}^{(n)}$ and $\alpha$, $K:=|I(\bs{\beta})|\geq \lfloor (n+1)/2\rfloor$.
\vs
\noin
\tb{Proof}: By the definition of MAD (the median of the absolute deviations to the center (median)), it is readily seen that
\bee
|I(\bs{\beta})|&=&\sum_{i=1}^{n}\mathds{1}\left( \frac{|r_i-m(\bs{Z}^{(n)},\bs{\beta})|}{\sigma(\bs{Z}^{(n)},\bs{\beta})}\leq \alpha\right)\\[2ex]
&\geq& \sum_{i=1}^{n}\mathds{1}\left( \frac{|r_i-m(\bs{Z}^{(n)},\bs{\beta})|}{\sigma(\bs{Z}^{(n)},\bs{\beta})}\leq 1\right)=\lfloor (n+1)/2\rfloor,
\ene
This completes the proof.
\hfill \pend
\vs
The lemma implies that the RHS of (\ref{objective.eqn}) sums a majority of squared residuals.
\vs
\noindent
\tb{Properties of the objective function}\vs
Write $D_i:=D(r_i, \bs{\beta})={|r_i-m(\bs{Z}^{(n)},\bs{\beta})|}\big/{\sigma(\bs{Z}^{(n)},\bs{\beta})}$ for a given $\mb{Z}^{(n)}$ and $\bs{\beta}$.
Let $i_1,\cdots, i_{K}$ in $I(\bs{\beta})$ such that $D_{i_1}\leq D_{i_2}\cdots\leq D_{i_K}$ (i.e. ordered depth values of residuals).
Both $i_j$ and $D_{i_j}$ clearly depend on $\bs{\beta}$ and $\mb{Z}^{(n)}$.\vs
Generally, the inequalities between $D_i$'s cannot be strict unless we assume that $r:=y-\bs{w}'\bs{\beta}$ has a density for any $\bs{\beta}\in \R^p$.
In the latter case, the strict inequalities hold almost surely (a.s.), i.e., $D_{i_1}< D_{i_2}\cdots< D_{i_K}$ (a.s.). Define for any $\bs{\beta}^1 \in\R^p$ and a given $\mb{Z}^{(n)}$
\be
R_{\bs{\beta}^1}=\{\bs{\beta}\in \R^p: I(\bs{\beta})=I(\bs{\beta}^1), D_{i_1}(\bs{\beta})< D_{i_2}(\bs{\beta})\cdots< D_{i_K}(\bs{\beta})\}. \label{region.eqn}
\ee
If $y-\bs{w}'\bs{\beta}$ has a density at $\bs{\beta}^1\in \R^p$, then $R_{\bs{\beta}^1} \neq \emptyset$ (a.s.).
There are at most finitely many $R_{\bs{\beta}^k}$s, $\bs{\beta}^k \in \R^p$, $1\leq k\leq L\leq {n \choose \lfloor (n+1)/2\rfloor }$ such that $\cup_{k=1}^L \overline R_{\bs{\beta}^k}=\R^d$, where $R_{\bs{\beta}^k}$ is defined similarly to (\ref{region.eqn}) and $\overline {A}$ stands for the closure of the set $A$.
For any $\bs{\beta} \in \R^p$, either there is $R_{\bs{\eta}}$ and
$\bs{\beta} \in R_{\bs{\eta}}$ or there is $R_{\bs{\xi}}$, such that $\bs{\beta} \not\in R_{\bs{\eta}}\cup R_{\bs{\xi}}$ and $\bs{\beta}\in \overline{R}_{\bs{\eta}}\cap \overline{R}_{\bs{\xi}}$. In the latter case, there are $i_k, i_l\in I(\bs{\beta})$ $i_k\neq i_l$,
such that $D_{i_k}=D_{i_l}$.
\vs
For a given sample $\bs{Z}^{(n)}$,
write $Q^n(\bs{\beta}) $ for $Q(\bs{Z}^{(n)}, \bs{\beta}, \alpha)$ and $B(\bs{\eta}, \delta)$ for an open ball in $\R^p$ centered at $\bs{\eta}$ with a radius $\delta>0$,
and $\mathds{1}_i$ for $\mathds{1}\left( {|y_i-\bs{w'}_i\bs{\beta}-m_n(\bs{\beta})|}\big/{\sigma_n(\bs{\beta})}\leq \alpha\right)$.  Note that $\mathds{1}_i$ depends on $\bs{\beta}$.
 Let $\mb{Y}_n=(y_1, \cdots, y_n)'$ and 
 $\bs{M}_n:=\bs{M}(\mb{Y}_n, \mb{X}_n, \bs{\beta}, \alpha)=\sum_{i=1}^n\bs{w}_i\bs{w}'_i\mathds{1}_i=\sum_{i\in I(\bs{\beta})} \bs{w}_i\bs{w}'_i$.
 We have 
\vs\noin
\tb{Lemma 2.2}\vs
\tb{(i)} For a given $\mb{Z}^{(n)}$ and $\alpha$, for any $1\leq k\leq L$ and any $\bs{\eta} \in R_{\bs{\beta}^k}$, there exists a $B(\bs{\eta}, \delta)$ such that for any $\bs{\beta} \in B(\bs{\eta},\delta)$, $\bs{\beta} \in R_{\bs{\beta}^k}$, i.e.,
$$Q^n(\bs{\beta})=\sum_{i\in I(\bs{\beta}^k)}r^2_i,$$\vs
\tb{(ii)} For any $1\leq k\leq L$, $R_{\bs{\beta}^k}$ is open,\vs
\tb{(iii)} $Q^n(\bs{\beta})$ is continuous in $\bs{\beta}\in \R^p$,\vs
\tb{(iv)} Over each $R_{\bs{\beta}^k}$, $1\leq k\leq L$, $Q^n(\bs{\beta})$ is twice differentiable and convex, and strictly convex if the rank of $\bs{X}_n$ is $p$.
\vs
\noin
\tb{Proof}:~ See the Appendix. \hfill \pend
\vs
\vs
\noindent
\tb{Remarks 2.1}\vs
\tb{(i)} By discussions above and Lemma 2.2, we see that the graph of $Q^n(\bs{\beta})$ is composed of components that are the graph of the quadratic function of the sum of squared residuals over each region $R_{\bs{\beta}^k}$, $1\leq k\leq L$.
\vs
\tb{(ii)} The continuity deduced from $Q^n(\bs{\beta})$ being the sum of some squared residuals without (i) of Lemma 2.2  might not be flawless. The unified expression for $Q^n(\bs{\beta})$ around the small neighborhood of $\bs{\beta}$ such as the one given in (i) of the Lemma 2.2 is indispensable.
\hfill \pend

\subsection{Existence, uniqueness and equivariance}
\noindent
 \tb{Theorem 2.1}  
\vs
\tb{(i)} $\widehat{\bs{\beta}}^n_{lst}$ exists and is the unique local minima of $Q^n(\bs{\beta})$ over $R_{\bs{\beta}^{k_0}}$ for some $k_0$ ($1\leq k_0\leq L$).
\vs
\tb{(ii)}
Over $R_{\bs{\beta}^{k_0}}$,  $\widehat{\bs{\beta}}^n_{lst}$  is the solution of the system of equations
\be
 \sum_{i=1}^{n}(y_i-\bs{w}'_i\bs{\beta})\bs{w}_i\mathds{1}_i 
  =\mb{0}, \label{estimation.eqn}
\ee
\vs
\tb{(iii)}
Over $R_{\bs{\beta}^{k_0}}$, the unique solution is
\be
\widehat{\bs{\beta}}^n_{lst}=\bs{M}_n(\mb{Y}_n, \mb{X}_n, \widehat{\bs{\beta}}^n_{lst}, \alpha)^{-1}\sum_{i \in I(\bs{\beta}^{k_0})}y_i\bs{w}_i
\label{lts-iterative.eqn}
\ee

\vs
\noin
\tb{Proof}:~ See the Appendix. \hfill \pend
\vs
Note that $\bs{X}_n$ has a full rank is
 sufficient for the matrix in the theorem to be invertible.
The existence could also be established as follows.
In the sequel, we will assume that\vs
\noindent
 \tb{(A0)} there is no vertical hyperplane which contains at least $\lfloor (n+1)/2\rfloor$ points of $\mb{Z}^{(n)}$.
  \vs
  \noin
 This holds true with probability one if $(\bs{x}', y)'$ has a joint density or holds if $\mb{Z}^{(n)}$ is \emph{in general position} (see Section 3 for  definition) (assume that $n>2p+1$ hereafter). 
\vs
\noin
\tb{Theorem 2.2} The minimizer $\widehat{\bs{\beta}}^n_{lst}$ of $Q(\bs{Z}^{(n)}, \bs{\beta}, \alpha)$ defined in (\ref{objective.eqn}) over $\bs{\beta}\in\R^p$ always
exists  for a given $\bs{Z}^{(n)}$ and $\alpha$ provided that \tb{(A0)} holds. 
\vs
\noin
\tb{Proof}:~ See the Appendix. \hfill \pend
\vs
\noindent
\tb{Equivariance}~
A regression estimator $\mb{T}$ is called \emph{regression, scale, and affine equivariant}
 if,  respectively (see page 116 of RL87) with $N=\{1,2, \cdots, n\}$
  \bee \mb{T}\left(\{(\bs{w}'_i, y_i+\bs{w}'_i \mb{b})', i\in N\}\right)&=&
\mb{T}\left(\{(\bs{w}'_i, y_i)', i\in N\}\right)+\mb{b}, ~\forall~ \mb{b}\in\R^p \label{regression.equi}\\
\mb{T}\left(\{(\bs{w}'_i, s y_i)', i\in N\}\right)&=&
s\mb{T}\left(\{(\bs{w}'_i, y_i)', i\in N\}\right), ~\forall~ s\in\R^1\\
\mb{T}\left(\{(A'\bs{w}_i)', y_i)', i\in N\}\right)&=&
A^{-1}\mb{T}\left(\{(\bs{w}'_i, y_i)', i\in N\}\right),~\forall~ \text{nonsingular}~ A\in \R^{ p\times p}
\ene
\vs
\vs
\noindent
\tb{Theorem 2.3}~ $\widehat{\bs{\beta}}^n_{lst}$ is regression, scale, and affine equivariant.
\vs
\noindent
\tb{Proof}: We have the identities
\bee
y_i+\bs{w}'_i\mb{b}-\bs{w}'_i(\bs{\beta}+\mb{b})&=&y_i-\bs{w}'_i\bs{\beta}, ~\forall~ \mb{b}\in\R^p  \\[1ex]
sy_i-\bs{w}'_i(s\bs{\beta})&=&s(y_i-\bs{w}'_i\bs{\beta}),~\forall~ s\in\R^1 \\[1ex]
y_i-(A'\bs{w}_i)' A^{-1}\bs{\beta}&=&y_i-\bs{w}'_i\bs{\beta}, ~\forall~ \text{nonsingular}~ A\in \R^{ p\times p}.
\ene
The desired result follows by these identities and the (regression, scale, and affine) invariance (see page 148 of Zuo (2021a) for definition) of $\frac{|r_i-m(\bs{Z}^{(n)},~\bs{\beta})|}{\sigma(\bs{Z}^{(n)},~\bs{\beta})}$. \hfill \pend

\vs
\section{Robustness of LST}
\subsection{Finite sample breakdown point}
As an alternative to the least-squares, is the LST estimator more robust?  The most prevailing quantitative measure of global robustness of any location or regression estimators in the finite sample practice is the
\emph{finite sample breakdown point} (FSBP), introduced by Huber and Donoho (1983) (DH83). 
\medskip

Roughly speaking, the FSBP is the minimum fraction of `bad' (or contaminated) data that the estimator can be affected to an arbitrarily large extent. For example, in the context of estimating the center of  a data set,
the sample mean has a breakdown point of $1/n$ (or $0\%$), because even one bad observation can change the mean
by an arbitrary amount; in contrast, the median has a breakdown point of $\lfloor(n+1)/2\rfloor/n$ (or $50\%$). 
\vs

\noindent
\textbf{Definition 3.1} [DH83] ~
The finite sample \emph{replacement breakdown point} (RBP) of a regression estimator $\mb{T}$ at the given sample
$\mb{Z}^{(n)}=\{Z_1,Z_2,\cdots,Z_n\}$, where $Z_i:=(\bs{x}_i', y_i)'$, is defined  as
\begin{equation}
\text{RBP}(\mb{T},\mb{Z}^{(n)}) = \min_{1\le m\le n}\bigg\{\frac{m}{n}: \sup_{\mb{Z}_m^{(n)}}\|\mb{T}(\mb{Z}_m^{(n)})- \mb{T}(\mb{Z}^{(n)})\| =\infty\bigg\},
\end{equation}
where $\mb{Z}_m^{(n)}$
denotes an arbitrary contaminated sample by replacing $m$ original sample points in $\mb{Z}^{(n)}$ with arbitrary points in $\R^{p}$.
 Namely, the RBP of an estimator is the minimum replacement fraction that could drive
the estimator beyond any bound.  It turns out that both $L_1$ (least absolute deviations) and $L_2$ (least squares) estimators have RBP $1/n$ (or $0\%$), the lowest possible value whereas LTS can have $( \lfloor( n - p ) / 2 \rfloor + 1)/n$ (or $50\%$), the highest possible value for any regression equivariant estimators (see pages 124-125 of RL87).
\vskip 3mm
We shall say  $\mb{Z}^{(n)}$  is\emph{ in general position}
when any $p$ of observations in $\mb{Z}^{(n)}$ gives a unique determination of $\bs{\beta}$.
In other words, any (p-1) dimensional subspace of the space $(\bs{x'}, y)'$ contains at most p observations of
$\mb{Z}^{(n)}$.
When the observations come from continuous distributions, the event ($\mb{Z}^{(n)}$ being in general position) happens with probability one.\vs
\noin
\tb{Theorem 3.1} For $\widehat{\bs{\beta}}^n_{lst}$ defined in (\ref{lst.eqn}) and $\mb{Z}^{(n)}$ in general position, we have 
\be
\text{RBP}(\widehat{\bs{\beta}}^n_{lst}, \mb{Z}^{(n)})=\left\{
\begin{array}{ll}
\lfloor (n+1)/2\rfloor\big/n, & \text{if $p=1$,}\\[1ex]
(\lfloor{n}/{2}\rfloor-p+2)\big/n,& \text{if $p>1$.}\\
\end{array}
\right. \label{T*-bp.eqn}
\ee
\vskip 3mm

\noindent
\textbf{Proof:}~~ See the Appendix. \hfill \pend
\vs
\noin
\tb{Remarks 3.1}

\vs
    \tb{(I)}  The assumption that $\mb{Z}^{(n)}$ is in general position seems to play a central role in the proof. But actually, one can drop it and introduce an index: $c(\mb{Z}^{(n)})$ (which is the maximum number of observations from $\mb{Z}^{(n)}$ contained in any $(p-1)$ dimensional subspace/hyperplane) to replace $p$ in the derivation of the proof and the final RBP result (when $p>1$).
\vs
\tb{(II)} Asymptotically speaking (i.e. as $n\to \infty$), $\widehat{\bs{\beta}}^n_{lst}$ has the best possible asymptotic breakdown point (ABP) $50\%$, the same as that of the  LTS. The RBP of $\widehat{\bs{\beta}}^n_{lst}$,  albeit very high (indeed as high as that of the LMS), is slightly less than that of LTS (with the best choice of $h$). However, it can  be improved to attain the best possible value
 if one modifies $\alpha $ so that it is the $h$th quantile of the $n$ depths of residuals with
 $h=\lfloor n/2\rfloor+\lfloor(p+1)/2\rfloor$ to include exact $h$ squares of residuals in the sum of the RHS of (\ref{objective.eqn}).  \hfill \pend
\vs
\subsection{Influence function}
Throughout $F_{\mb{z}}$ stands for the distribution of random vector $\mb{z}$ unless otherwise stated.
Write $F_{(\bs{x'}, y)}$ for the joint distribution of $\bs{x}'$ and $y$ in (\ref{model.eqn}), $r:=r(F_{(\bs{x'}, y)}, \bs{\beta})=y-(1,\bs{x}')\bs{\beta}:=y-\bs{w'}\bs{\beta}$. 
\begin{align}
  m:=&m(F_{(\bs{x'}, y)}, \bs{\beta})=\mbox{Med}(F_r),\nonumber\\
  \sigma:=&\sigma(F_{(\bs{x'}, y)}, \bs{\beta})=\mbox{MAD}(F_r),\nonumber
  \end{align}
hereafter we assume that $m$ and $\sigma$ exist uniquely.
The population counterparts of (\ref{objective.eqn}) and (\ref{lst.eqn}) are respectively:
\begin{align}
Q(F_{(\bs{x'}, y)},\bs{\beta}, \alpha):&=\int(y-\bs{w}'\bs{\beta})^2\mathds{1}\left( \frac{|y-\bs{w}'\bs{\beta}-m|}{\sigma}\leq \alpha\right)dF_{(\bs{x'}, y)}, \label{Q.eqn}\\[1ex]
\bs{\beta}_{lst}(F_{(\bs{x'}, y)}, \alpha) :&=\arg\min_{\bs{\beta}\in \R^p}Q(F_{(\bs{x'}, y)}, \bs{\beta}, \alpha). \label{lst-def.eqn}
\end{align}

\vs
RBP measures the global robustness of an estimator at finite sample practice. To investigate the local robustness at the population setting, one can use the influence function approach (see Hampel, et al. 1986 (HRRS86)), which depicts the local robustness of a functional with an 
infinitesimal point-mass contamination at a single point $\bs{z}\in \R^{p}$.\vs

For a given distribution $F$ defined on $\R^{p}$ 
 and an $\ep > 0$, the version of $F$ contaminated by an $\ep$ amount of an \emph{arbitrary distribution} $G $ on $\R^{p}$ is
denoted by $F(\ep, G) = (1 -\ep)F + \ep G$ (an $\ep$ amount deviation from the assumed $F$). Hereafter it is assumed that $\ep < 1/2$, otherwise $F(\ep, G)=G((1-\ep), F)$, and one can't distinguish which one is contaminated by which one.\vs
\noindent
\tb{Definition 3.2}~[HRRS86]~
The \emph{influence function} (IF) of a functional $T$ at a given point $\bs{z} \in \R^{p}$ for a given $F$ is defined
as
\be
\text{IF}(\bs{z}; \bs{T},F) = \lim_ {\ep\to 0^+} \frac{\bs{T} (F (\ep, \delta_{\bs{z}})) - \bs{T} (F )}{\ep},\label{if.eqn} \ee
where $\delta_{\bs{z}}$ is the point-mass probability measure at $\bs{z} \in \R^{p}$.

\vs
The function $\text{IF}(\bs{z}; \bs{T},F)$ describes the relative effect (influence) on $\bs{T}$ of an
infinitesimal point-mass contamination at $\bs{x}$ and measures the local robustness
of $\bs{T}$.

\vs It is desirable that a regression estimating functional has a bounded influence function. 
This, however, does not hold for an arbitrary regression estimating functional (such as the classical least squares functional). Now we investigate this for the functional of the least sum of squares of trimmed residuals, $\bs{\beta}_{lst}(F_{(\bs{x'}, y)}, \alpha)$.
 Put
\begin{align}
 F_{\varepsilon}(\mb{z}):=&F(\varepsilon, \delta_{\mb{z}})= (1-\varepsilon) F_{(\bs{x'}, y)}+\varepsilon \delta_{\mb{z}}, \nonumber\\
m_{\varepsilon}(\mb{z}):=&m(F_{\varepsilon}(\mb{z}),\bs{\beta})=\mbox{Med}(F_{R_{\varepsilon}(\mb{z})}),\nonumber\\ \sigma_{\varepsilon}(\mb{z}):=&\sigma(F_{\varepsilon}(\mb{z}),\bs{\beta})=\mbox{MAD}(F_{R_{\varepsilon}(\mb{z})}),\nonumber
\end{align}
where  $R_{\varepsilon}(\mb{z})=r(F_{\varepsilon}(\mb{z}), \bs{\beta})=t-(1,\mb{s}')\bs{\beta}$, and $F_{\varepsilon}(\mb{z})$ with $\mb{z}=(\mb{s}', t)' \in \R^{p}$, $\mb{s}\in \R^{p-1}$, and $t\in \R^1$. Here after we assume that $m_{\varepsilon}(\mb{z})$ and $\sigma_{\varepsilon}(\mb{z})$ are uniquely exist.
The versions of (\ref{Q.eqn}) and (\ref{lst-def.eqn}) at the contaminated distribution $F_{\varepsilon}(\mb{z})$ are respectively
\begin{align}
Q(F_{\varepsilon}(\mb{z}),\bs{\beta}, \alpha):&=\int(t-(1,\mb{s}')\bs{\beta})^2\mathds{1}\left( \frac{|(t-(1,\mb{s}')\bs{\beta})-m_{\varepsilon}(\mb{z})|}{\sigma_{\varepsilon}(\mb{z})}\leq \alpha\right)dF_{\varepsilon}(\mb{z}),\label{Q-contaminated.eqn}\\[1ex]
\bs{\beta}_{lst}(F_{\varepsilon}(\mb{z}), \alpha) :&=\arg\min_{\bs{\beta}\in \R^p}Q(F_{\varepsilon}(\mb{z}), \bs{\beta}, \alpha).
\end{align}

\noin
\tb{Lemma 3.1} $\bs{\beta}_{lst}:= \bs{\beta}_{lst}(F_{(\bs{x'}, y)}, \alpha)$ is regression, scale, and affine equivariant (see Zuo (2021a) for definition).
\vs
\noindent
\tb{Proof}: It is trivial (analogous to that of  Theorem 2.3). \hfill \pend
\vs
To investigate the influence function of $\bs{\beta}_{lst}$ especially the consistency of its sample version in the next section, we first need to establish its existence and uniqueness. We need assumptions:
\tb{(A1)} $y$ has a density, and
\tb{(A2)} the distribution $F_r$ with $r=y-\bs{w}'\bs{\beta}$ is non-flat around $m=\mbox{Med}(F_r)$ and $\sigma=\mbox{MAD}(F_r)$ for any $\bs{\beta} \in \R^p$.
\vs
Write $Q(\bs{\beta})$ for $Q(F_{(\bs{x}', y)}, \bs{\beta}, \alpha)$ in (\ref{Q.eqn}). We have a population counterpart of Lemma 2.2.
\vs
\noindent
\tb{Lemma 3.2} Assume \tb{(A1)-(A2)} hold. 
 Then $Q(\bs{\beta})$\vs
\tb{(i)} 
is continuous in $\bs{\beta} \in \R^p$;
\vs
\tb{(ii)} 
is twice differentiable  in $\bs{\beta} \in \R^p$ with \[{\partial^2 Q(\bs{\beta})}\big/{\partial \bs{\beta}^2}=2E\bs{w}\bs{w}' \mathds{1}\left({|y-\bs{w}'\bs{\beta}-m|}\big/{\sigma}\leq \alpha\right);\]
\vs
\tb{(iii)} 
is  convex in $\bs{\beta} \in \R^p$ and strictly convex if $E \bs{w}\bs{w}' \mathds{1}\left( {|y-\bs{w}'\bs{\beta}-m|}\big/{\sigma}\leq \alpha\right)$ is invertible .

\noindent
\tb{Proof:} See the Appendix. \hfill \pend
 \vs

 \noin
 \tb{Theorem 3.2} Under the assumptions of Lemma 3.2 and assume that $m(F_{\varepsilon}(\mb{z}),\bs{\beta})$ and $\sigma(F_{\varepsilon}(\mb{z}),\bs{\beta})$ are continuous in $\bs{\beta}$ around a small neighborhood of $\bs{\beta}_{lst}((F_{\varepsilon}(\mb{z}), \alpha)$. Write $\bs{v}'=(1, \bs{s}')$  and let $\bs{u}$ be the random variable with CDF $F_{\varepsilon}(\bs{z})$.  We have\vs
 \tb{(i)} $\bs{\beta}_{lts}(F_{(\bs{x}', y)}, \alpha)$  and $\bs{\beta}_{lts}(F_{\varepsilon}(\mb{z}), \alpha)$  exist.
 \vs
 \tb{(ii)} Furthermore, they are the solution of system of equations,  respectively
 \begin{align}
 \int(y-\bs{w}'\bs{\beta})\bs{w}\mathds{1}\left( {|y-\bs{w}'\bs{\beta}-m|}\big/{\sigma}\leq \alpha\right) dF_{(\bs{x}', y)}(\bs{x}, y)&=\mb{0},  \label{lst-estimation.eqn} \\[1ex]
 \int(t-\bs{v}'\bs{\beta})\bs{v}\mathds{1}\left( {|(t-\bs{v}'\bs{\beta})-m_{\varepsilon}(\mb{z})|}\big/{\sigma_{\varepsilon}(\mb{z})}\leq \alpha\right) dF_{\mb{u}}(\mb{s}, t)&=\mb{0}. \label{lts-contamination-estimation.eqn}
 \end{align}

 \tb{(iii)} $\bs{\beta}_{lts}(F_{(\bs{x}', y)}, \alpha)$  and $\bs{\beta}_{lts}(F_{\varepsilon}(\mb{z}), \alpha)$ are unique
  provided that
 \begin{align}
  \int \bs{w}\bs{w}'\mathds{1}\left( {|y-\bs{w}'\bs{\beta}-m|}\big/{\sigma}\leq \alpha\right)dF_{(\bs{x}', y)}(\bs{x}, y)&, \label{uniqueness-lts.eqn}\\[1ex]
  \int \bs{v}\bs{v}'\mathds{1}\left( {|(t-\bs{v}')\bs{\beta})-m_{\varepsilon}(\mb{z})|}\big/{\sigma_{\varepsilon}(\mb{z})}\leq \alpha\right) dF_{\mb{u}}(\mb{s}, t)&
  \end{align}
  \hspace*{15mm}{are respectively  invertible.}
 \vs
 \noindent
\tb{Proof:} See the Appendix. \hfill \pend
\vs
\noindent
\tb{Theorem 3.3} If assumptions in theorem 3.2 hold,
then for any $\mb{z}_0:=(\mb{s'}_0, t_0) \in \R^{p}$, we have that 
\[
\dot{\bs{\beta}}_{lst} (\mb{z}_0, F_{(\bs{x'}, y)})=\left\{
\begin{array}{ll}
\mb{0},& \hspace*{-35mm}\mbox{if}~ t_0-(1, \mb{s}'_0)\bs{\beta}_{lst} \not \in [m(\bs{\beta}_{lst})-\alpha\sigma(\bs{\beta}_{lst}),~~ m(\bs{\beta}_{lst})+\alpha\sigma(\bs{\beta}_{lst})],  \\[1ex]
(t_0-(1, \mb{s}'_0)\bs{\beta}_{lst}) M^{-1}(1, \mb{s}'_0)',& \mbox{otherwise},
\end{array}
\right.
\]
\vs
\noin
where $\dot{\bs{\beta}}_{lst} (\mb{z}_0, F_{(\bs{x'}, y)})$ stands for the $\mbox{IF}(z_0; \bs{\beta}_{lst}, F_{(\bs{x'}, y)})$ and $M^{-1}$ stands for the inverse of the matrix $E\left(\bs{w}\bs{w'}\mathds{1}\left( {|r(\bs{\beta})-m(F_{r(\bs{\beta})})|}\big/{\sigma(F_{r(\bs{\beta})})}\leq \alpha\right)\right)$ with $\bs{\beta}=\bs{\beta}_{lst}$.
\vs
\noindent
\tb{Proof}:~~ See the Appendix. \hfill \pend
\vs
\noin
\vs
\noindent
\tb{Remarks 3.2} ~~ see the Appendix. \hfill \pend
\vs
Overall, we see that LST is globally robust with the best possible ABP of $50\%$ and robust locally against point-mass contamination when there are vertical and bad leverage outliers.
\vs
Besides robustness,
one wonders: does the $\bs{\beta}_{lst}(F_{(\bs{x}',y)},\alpha)$
really catch the true parameter (i.e. is it Fisher consistent)? And how fast does the sample $\bs{\beta}_{lst}(Z^{(n)})$ converge to $\bs{\beta}_{lst}$ (or the true parameter $\bs{\beta}_0$) (i.e. strong or root-n consistency)?  We answer these questions next.

\vs
\section{Consistency}
\subsection{Fisher Consistency}
Before establishing the strong or root-n consistency, we like to first show that the population version of LST, $\bs{\beta}_{lst}(F_{(\bs{x'}, y)}, \alpha)$, is consistent with (identical to) the true unknown parameter $\bs{\beta}_0$ under some assumptions - which is called Fisher consistency of the estimation functional.
To that end, let us first recall our general model:
\be
y=(1,\bs{x}')\bs{\beta}_0+e, \label{model-2.eqn}
\ee
with its sample version given in model (\ref{model.eqn}). In addition to the assumptions given in Theorem 3.2 for the existence and uniqueness of $\bs{\beta}_{lst}$, we need one more assumption:
\vs
\tb{(A3)} $\bs{x}$ and $e$ are independent and $E_{{(\bs{x'}, y)}}\left( e \mathds{1}\left( {|e-m({F_e})|}\big/{\sigma(F_e)}\leq \alpha\right)\right)=0$ , where $F_e$ is the distribution of r.v. $e$. Hereafter we assume that $m(F_e)$ and $\sigma(F_e)$ exist uniquely.
\vs
The independence assumption between $\bs{x}$ and $e$ is typical in the traditional regression analysis. However, one can drop it here by modifying the integration appropriately (see the proof below), and it is unnecessary if $\bs{x}$ is a non-random covariate (carrier).
 The assumption that integration equals to zero is very mild, and it automatically holds under the common assumption that the $e$ is symmetric with respect to $0$ (that is, $e\stackrel{d}=-e$).
We have
\vs
\noin
\tb{Theorem 4.1} Under assumptions \tb{(A1)-(A3)},  $\bs{\beta}_{lst}(F_{(\bs{x'}, y)}, \alpha)=\bs{\beta}_0$ (i.e. it is Fisher consistent). 
\vs
\noindent
\tb{Proof}: Notice that $y-\bs{w'}\bs{\beta}=\bs{w'}(\bs{\beta}_0-\bs{\beta})+e$. This in conjunction with equation (\ref{lst-estimation.eqn}) yields,
$$
\int(\bs{w'}(\bs{\beta}_0-\bs{\beta})+e)\bs{w}\mathds{1}\left( {|(\bs{w'}(\bs{\beta}_0-\bs{\beta})+e)-m|}\big/{\sigma}\leq \alpha\right)dF_{(\bs{x'}, y)}=\mb{0},
$$
 one  sees that $\bs{\beta}=\bs{\beta}_0$ indeed is one solution of the equation system by virtue of \tb{(A3)}. In light of Theorem 3.2 and the uniqueness of the solution, the desired result follows.
\hfill \pend
\vs
\subsection{Strong consistency}
\vs
To establish the strong consistency of $\widehat{\bs{\beta}}_{lst}(\mb{Z}^{(n)}, \alpha)$ for the $\bs{\beta}_{lst}(F_{(\bs{x'}, y)}, \alpha)$, for notation simplicity we write $\widehat{\bs{\beta}}_{lst}(F^n_{\mb{Z}}):=\widehat{\bs{\beta}}_{lst}(\mb{Z}^{(n)}, \alpha)$ and $\bs{\beta}_{lst}(F_{\mb{Z}}):= \bs{\beta}_{lst}(F_{(\bs{x'}, y)}, \alpha)$
and $Q(F^n_\mb{Z}, \bs{\beta}):=Q(\mb{Z}^{(n)}, \bs{\beta}, \alpha)$ and $Q(F_\mb{Z}, \bs{\beta}):=Q(F_{(\bs{x}', y)}, \bs{\beta}, \alpha)$, 
 where $F^n_\mb{Z}$ is the sample version of $F_\mb{Z}:=F_{(\bs{x'}, y)}$, corresponding to $\mb{Z}^{(n)}$ and $\alpha$ is suppressed.\vs
We will follow the approach in Zuo (2020) and treat the problem in a more general setting.
To that end, we introduce the regression depth functions $D(F^n_\mb{Z}, \bs{\beta})=(1+Q(F^n_\mb{Z}, \bs{\beta}))^{-1}$ and $D(F_{\mb{Z}}, \bs{\beta})=(1+Q(F_\mb{Z}, \bs{\beta}))^{-1}$ (see page 144 of Zuo (2021a) for the objective function approach). The original  minimization problem becomes a maximization problem.\vs
Let $M_n$  be stochastic processes indexed by a metric space $\Theta$ of $\bs{\theta}$, and $M\!\!:$ $\Theta \to\R$ be a deterministic function of $\bs{\theta}$ which attains its maximum at a point $\bs{\theta}_0$.\vs
 The sufficient conditions  for the consistency of this type of problem were given in Van Der Vaart (1998) (VDV98) and Van Der Vaart and Wellner (1996) (VW96), they are:\vs
\begin{itemize}
\item[]\tb{C1:} $\sup_{\bs{\theta}\in\Theta}|M_n(\bs{\theta})-M(\bs{\theta})|=o_p(1)$;
\vskip 3mm
\item[]\tb{C2:} $\sup_{~\{\bs{\theta}:~d(\bs{\theta},\bs{\theta_0})\geq \delta\}} M(\bs{\theta})< M(\bs{\theta_0})$, for any $\delta>0$ and the metric $d$ on $\Theta$;
\vskip 3mm
\hspace*{-10mm} Then any sequence $\bs{\theta}_n$ is consistent for $\bs{\theta}_0$ providing  that it satisfies
\item[]\tb{C3:} $M_n(\bs{\theta_n})\geq M_n(\bs{\theta_0})-o_p(1)$.
\vskip 3mm

\end{itemize}

\vs
\noin
\tb{Lemma 4.1} [Th. 5.7, VDV98] If \tb{C1} and \tb{C2} hold, then any $\bs{\theta}_n$ satisfying \tb{C3} is consistent for $\bs{\theta}_0$. \hfill \pend

\vs
\noin
\tb{Remarks 4.1}
\vs
\tb{(I)} \tb{C1} requires that the $M_n(\bs{\theta})$  converges   to $M(\bs{\theta})$ in probability  uniformly in $\bs{\theta}$. For the depth process D$(F^n_{\mb{Z}}, \bs{\beta})$ and D$(F_{\mb{Z}}, \bs{\beta})$, it holds true (the convergence here is almost surely (a.s.)
     and uniformly in $\bs{\beta}$ as shown in Lemma 4.2 below).
\vs
\tb{(II)} \tb{C2} essentially demands that the unique maximizer $\bs{\theta_0}$ is well separated. This holds true for D$(F_{\mb{Z}}, \bs{\beta})$ as shown in Lemma 4.3 below.
\vs
\tb{(III)} \tb{C3} asks that $\bs{\theta_n}$
 is very close to $\bs{\theta}_0$ in the sense that the difference of images of the two at $M_n$ is within $o_p(1)$.
In Kim and Pollard (1990) (KP90) and VW96 a stronger version of \tb{C3} is required:
\vs
$\displaystyle \mbox{\tb{C3}*}: ~~ M_n(\bs{\theta_n})\geq \sup_{\theta\in\Theta} M_n(\bs{\theta})-o_p(1),$\vs
 which implies \tb{C3}. This strong version mandates that $\bs{\theta_n}$  \emph{nearly} maximizes $M_n(\bs{\theta})$.
Our maximum regression depth estimator $\widehat{\bs{\beta}}_{lst}(F^n_{Z}, \alpha)(:=\bs{\theta_n})$ is defined to be the maximizer of $M_n(\bs{\theta}):=D(F^n_{\mb{Z}}, \bs{\beta})$, hence \tb{C3}* (and thus \tb{C3}) holds automatically. \hfill \pend
\vs
In light of above, we have
\vs
\noin
\tb{Corollary 4.1} $\widehat{\bs{\beta}}_{lst}(F^n_{\mb{Z}})$ induced from D$(F^n_{\mb{Z}}, \bs{\beta})$ (or $Q(F^n_{\mb{Z}}, \bs{\beta})) $ is consistent for $\bs{\beta}_{lst}(F_{\mb{Z}})$.  \hfill \pend
\vs
But, we can have more.
\vs
\noindent
\tb{Theorem 4.2} Under assumptions \tb{(A1) -(A3)}, $\widehat{\bs{\beta}}_{lst}(F^n_{\mb{Z}})$
 is \emph{strongly} consistent for $\bs{\beta}_{lst}(F_{\mb{Z}})$ (i.e., $\widehat{\bs{\beta}}^n_{lst}-\bs{\beta}_{lst}=o(1)$ a.s.).
\vs
\noindent
\tb{Proof:} 
The proof for the consistency of Lemma 4.1 could be easily extended to the strong consistency with a strengthened version  of \tb{C1}

 \vs
$\displaystyle ~~~~\mbox{\tb{C1}*:}~~\sup_{\bs{\theta}\in\Theta}|M_n(\bs{\theta})-M(\bs{\theta})|=o(1)$, a.s.,\vs
In the light of the proof of Lemma 4.1, we need only  verify the sufficient conditions \tb{C1}* and \tb{C2-C3}.  By (III) of Remark 4.1,
\tb{C3} holds automatically, so  we need to verify \tb{C1}* and \tb{C2}.
\tb{C1}* will be given in Lemma 4.2. So the only item left is to verify  \tb{C2} for
D$(F_{Z}, \bs{\beta})$ which is guaranteed by Lemma 4.3 below. 
\hfill \pend
\vs
Based on the proofs of Theorems 2.2 and 3.2 and in light of Theorem 4.1, under assumptions \tb{(A0)-(A3)}, we assume without loss of generality (w.l.o.g.) that $\widehat{\bs{\beta}}_{lst}(F^n_{\mb{Z}})\in B(\bs{\beta}_0, r)$ and ${\bs{\beta}}_{lst}(F_{\mb{Z}})\in B(\bs{\beta}_0, r)$,
where $B(\bs{\beta}_0, r)$ is a ball centered at $\bs{\beta}_0$ with radius $r$ which is large enough.  Now $B(\bs{\beta}_0, r)$ can serve, w.l.o.g., as out parameter space $\Theta$ of $\bs{\beta}$ in the sequel.
\vs
\noindent
\tb{Lemma 4.2} Under assumption \tb{(A2)}, (a) $\sup_{\bs{\beta}\in\Theta}|Q(F^n_{\mb{Z}},\bs{\beta})-Q(F_{\mb{Z}},\bs{\beta})|=o(1)$, a.s. and (b) $\sup_{\bs{\beta}\in\Theta}|D(F^n_{\mb{Z}},\bs{\beta})-D(F_{\mb{Z}},\bs{\beta})|=o(1)$, a.s..
\vs
\noindent
\tb{Proof}:~~ See the Appendix. \hfill \pend

\vs
\noindent
\tb{Lemma 4.3} Assume that a regression (or location) depth function $D(\bs{\beta}; F_{\mb{Z}})$ is continuous in $\bs{\beta}$ and $\bs{\beta} \in \Theta$ is bounded. 
 Let $\bs{\eta}\in\Theta$ be the unique point with $\bs{\eta}=\arg\max_{\bs{\beta}\in\Theta}D(\bs{\beta}; F_{\mb{Z}})$ and $D(\bs{\eta}; F_{\mb{Z}})>0$.
Then for any $\ep>0$, $\sup_{\bs{\beta}\in N^c_{\ep}(\bs{\eta})}D(\bs{\beta}; F_{\mb{Z}}) < D(\bs{\eta}; F_{\mb{Z}})$, where $N^c_{\ep}(\bs{\eta})=\{\bs{\beta}\in\Theta: \|\bs{\beta}-\bs{\eta}\|\geq \ep\}$ and ``c" stands for ``complement" of a set.
\vs
\noin
\tb{Proof}:~ See the Appendix. \hfill \pend
\vs
\noindent
\tb{Remarks 4.2}
\vs
\tb{(I)} The approach utilizing a generalized Glivenko-Cantelli theorem over a class of functions with polynomial discrimination in the proof of lemma 4.2 is very powerful and applicable to many regression estimators to obtain the strong consistency result. It is certainly applicable to the least trimmed squares (LTS) estimator.
\vs
\tb{(II)} The consistency (not the strong version) of LTS has been obtained by V\'{i}\"{s}ek (2006a) using standard analysis (under many assumptions on non-random $\bs{x}_i$ and on the distribution of $e$)  which, of course, is difficult, lengthy (consumed an entire article), and tedious. The approach here is  different, concise and the estimator (LST) is, of course, different to LTS.
\hfill \pend
\vs
\vs
Consistency does not reveal the speed of convergence of sample $\widehat{\bs{\beta}}_{lst}(F^n_{\mb{Z}})$ to its population counterpart $\bs{\beta}_{lst}(F_{\mb{Z}})$. Standard speed of $O_p(1/\sqrt{n})$ is desirable and expected for $\widehat{\bs{\beta}}_{lst}(F^n_{\mb{Z}})$. We investigate this issue next.
\vs
\subsection{$\sqrt{n}$- consistency}

To establish the root-n consistency we need one more assumption:
\vs
\tb{(A4)}   E$(e)=0$ and E$(\bs{x}\bs{x'})$ exists.
\vs
 E$(e)=0$ is commonly required in the traditional regression analysis. The existence of covariance (and the mean) of $\bs{x}$ is sufficient for the existence of E$(\bs{x}\bs{x'})$. 
\vs
In the following, we will employ big $O$ and little $o$ notation for the vectors or matrices.\vs
\noin
\tb{Definition 4.1}  ~For a sequence of random vectors or matrices $\bs{X}_n$, we say

$\bs{X}_n=o_p(1)$ means $\|\bs{X}_n\|\stackrel{p} \to 0$;

$\bs{X}_n=O_p(1)$ means  $\|\bs{X}_n\|=O_p(1)$,

\noin
where norm of a matrix $A_{m\times n}$ is defined as $\|A\|:=\sup_{\bs{x}\not =0\in \R^n}{\|A\bs{x}\|_p}\big/{\|\bs{x}\|_p}$, $p$ could be $1, 2,~ \mbox{or} ~\infty$ (see page 82 of Boyd and Vandenberghe (2004) (BV04)).
\hfill \pend
\vs
\vs
\noindent
\tb{Theorem 4.3} Under assumptions \tb{(A0)}-\tb{(A4)}, $\widehat{\bs{\beta}}^n_{lst}-\bs{\beta}_{lst}=\widehat{\bs{\beta}}^n_{lst}-\bs{\beta}_0 =O_p(1/\sqrt{n})$.
\vs
\noin
\tb{Proof}:~ See the Appendix. \hfill \pend
\vs
\noindent
\tb{Remarks 4.4}
\vs
\tb{(I)} The root-n consistency of an $\arg\max$ estimator could be established by a general approach given in Sherman (1993, 1994) Theorem 1. With the depth process introduced in the section 4.2, we are unable to verify the second requirement in that theorem though.
\vs
\tb{(II)} The approach here for the root-n consistency of LST is analogous to what is given in V\'{i}\"{s}ek (2006b) for LTS. However, the latter is lengthy and costs a twenty-two pages article. 
\hfill \pend
\vs
\section{Asymptotic normality}
\vs
The root-n consistency above could be obtained as a by-product of the asymptotic normality which will be established in the following
via stochastic equicontinuity (see page 139 of Pollard 1984 (P84), or the supplementary of Zuo (2020)).

\vs
\emph{Stochastic equicontinuity} refers to a sequence of
stochastic processes $\{Z_n(t): t \in T\}$ whose shared index set $T$ comes equipped
with a semi metric $d(\cdot, \cdot)$.
\vs
\noin
\tb{Definition 5.1} [IIV. 1, Def. 2 of P84].  Call ${Z_n}$ stochastically equicontinuous at $t_0$  if for each $\eta > 0$
and $\epsilon > 0$ there exists a neighborhood $U$ of $t_0$ for which
\be
\limsup P\left(\sup_{U} |Z_n(t) - Z_n(t_0) | > \eta\right) < \epsilon.  \label{se.eqn}
\ee
\hfill~~~~~~~~~~~~~\pend
\vs
If ${\tau_n}$ is a sequence of random elements of $T$ that converges in probability
to $t_0$, then
\be
Z_n(\tau_n)-Z_n(t_0)\to 0\mbox{~in probability,}
\ee
because, with probability tending to one, $\tau_n$ will belong to each $U$.
The form above will be easier to apply, especially when behavior of a particular ${\tau_n}$ sequence is under investigation.
\vs
Suppose $\mathscr{F} = \{ f(\cdot, t): t\in T\}$, with $T$ a subset of $\R^k$, is a collection of
real, P-integrable functions on the set $S$ where $P$ (probability measure) lives.
Denote by $P_n$ the
empirical measure formed from $n$ independent observations on $P$, and define the empirical process $E_n$ as the signed measure $n^{1/2}(P_n - P)$. Define
\begin{align*}
F(t) &= P f(\cdot, t),\\
F_n(t) &= P_n f(\cdot, t).
\end{align*}
Suppose $f(\cdot, t)$ has a linear approximation near the $t_0$ at which $F(\cdot)$ takes
on its minimum value:
\be
 f(\cdot, t) = f(\cdot, t_0) + (t - t_0)'\nabla(\cdot) + |t - t_0|r(\cdot, t). \label{taylor.eqn}
\ee
For completeness set $r(\cdot, t_0) = 0$, where $\nabla$ (differential operator) is a vector of $k$ real functions on
$S$. We cite theorem 5 of IIV.1 of P84  (page 141) for the asymptotic normality of $\tau_n$.
\vs
\noindent
\tb{Lemma 5.1} .  Suppose $\{\tau_n\}$ is a sequence of random vectors converging in
probability to the value $t_0$ at which $F(\cdot)$ has its minimum. Define $r(\cdot, t)$ and the
vector of functions $\nabla(\cdot)$ by (\ref{taylor.eqn}). If
\bi
\item[(i)] $t_0$ is an interior point of the parameter set $T$; \vspace*{-2mm}
\item[(ii)] $F(\cdot)$ has a non-singular second derivative matrix $V$ at $t_0$;\vspace*{-2mm}
\item[(iii)] $F_n(\tau_n) = o_p(n^{-1}) + \inf_{t}F_n(t)$;\vspace*{-2mm}
\item[(iv)] the components of $\nabla(\cdot)$ all belong to $\mathscr{L}^2(P)$;\vspace*{-2mm}
\item[(v)] the sequence $\{E_{n}(\cdot, t)\}$ is stochastically equicontinuous at $t_0$ ;\vspace*{-2mm}
\ei
then
\[n^{1/2}(\tau_n - t_0) \stackrel{d} \longrightarrow  {\cal{N}}(O, V^{-1}[P(\nabla\nabla') - (P\nabla)(P\nabla)']V^{-1}).
\]
\hfill \pend
\vs
\noindent
\tb{Theorem 5.1} Assume that\vspace*{-2mm}
\bi
\item[(i)] the uniqueness assumptions for $\widehat{\bs{\beta}}^n_{lst}$ and $\bs{\beta}_{lst}$ in theorems 2.3 and 3.2 hold respectively;\vspace*{-2mm}
\item[(ii)] $P({x^2_i})$ exists; \vspace*{-2mm}
\ei
then
\[n^{1/2}(\widehat{\bs{\beta}}^n_{lst} - \bs{\beta}_{lst}) \stackrel{d} \longrightarrow  {\cal{N}}(O, V^{-1}[P(\nabla\nabla') - (P\nabla)(P\nabla)']V^{-1}),
\]
where $\bs{\beta}$ in $V$ and $\nabla$ is replaced by $\bs{\beta}_{lst}$ (which could be assumed to be zero).
\vs
\noindent
\tb{Proof}:~~ See the Appendix. \hfill \pend
\vs
Assume that $\bs{z}=(\bs{x}', y)'$ follows  elliptical distributions $E(g; \bs{\mu}, \bs{\Sigma})$ with density
\be
f_{\bs{z}}(\bs{x}', y)=\frac{g(((\bs{x}',y)'-\bs{\mu})'\bs{\Sigma}^{-1}((\bs{x}',y)'-\bs{\mu}))}{\sqrt{\det(\bs{\Sigma})}}, \label{ell.eqn}
\ee
where 
$\bs{\mu}\in \R^p$ and $\bs{\Sigma}$  a positive definite matrix of size $p$ which is proportional to the covariance matrix if the latter exists. We assume the function $g$ to have a strictly negative derivative, so that the $f_{\bs{z}}$ is unimodal.\vs
In light of Lemma 3.1 and under some transformations (see the Appendix), we can assume, w.l.o.g. that $(\bs{x}', y)$ follows an $E(g; \bs{0}, \bs{I}_{p\times p})$ (spherical) distribution and $\bs{I}_{p\times p}$ is the covariance matrix of $(\bs{x}', y)$ in the following.
\vs
\noindent
\tb{Corollary 5.1} Assume that
\bi
\item[(i)] assumptions of Theorem 4.1 hold;
\item[(ii)] $e\sim\mathcal{N}(0, \sigma^2)$ and $\bs{x}$ are independent.
\ei
Then
\bi
\item[(1)] $P\nabla=\bs{0}$ and $P(\nabla\nabla')=8\sigma^2 C \bs{I}_{p\times p}$,\\[1ex]
 with $C=\Gamma(1/2, 1)(\alpha c/\sigma)$ where $\Gamma(1/2, 1)(x)$ is the cumulative distribution function of random variable $\Gamma(\alpha,\beta)$ which has a pdf: $\frac{\beta^{\alpha}}{\Gamma(\alpha)} x^{\alpha-1}e^{-\beta x} $ and $c=\sigma\Phi^{-1}(3/4)$, where $\Phi (x) $ is the cumulative distribution function of $\mathcal{N}(0,1)$.
\item[(2)] $\mb{V}= 2C_1\bs{I}_{p\times p} $ with $C_1=2*\Phi(\alpha c/\sigma)-1$.
\item[(3)] $n^{1/2}(\widehat{\bs{\beta}}^n_{lst} - \bs{\beta}_{lst}) \stackrel{d} \longrightarrow  {\cal{N}}(\bs{0}, \frac{2C\sigma^2}{C_1^2}\bs{I}_{p \times p}).$
\ei
\vs
\noindent
\tb{Proof}: By Theorem 4.1 and Lemma 3.1, we can assume, w.l.o.g., that ${\bs{\beta}}_{lst}=\bs{\beta}_0=\bs{0}$. Utilizing the independence between $e$ and $\bs{x}$ and Theorem 4.4, a straightforward calculation leads to the results.
\hfill \pend

\section{Computation}
Now we address one of the most important topics on robust regression estimation, that is, the computation of the  estimator.
Unlike the LS estimator, which has an analytical formula for computation,
for the least sum of squared trimmed (LST) residuals  estimator, we do not have such a formula. The formula given in (\ref{lts-iterative.eqn}) can not serve our purpose (due to the circular dependency: the RHS depends on the LHS). For small sample size $n$ and dimension $p$, one can compute the LST exactly (the $L$ in Theorem 2.1 is not a big number), but that is not affordable for moderate sample size $n$ and dimension $p$.
That is, generally, 
 we have to appeal to approximate algorithms (AAs).\vs

Let us first recall our minimization problem and the quantity that needs to be minimized.
For a given data set $\mb{Z}^{(n)}=\{(\bs{x}'_i, y_i)'\}$ in $\R^{p}$ and an $\alpha\geq 1 $ and a $\bs{\beta} \in \R^p$, consider the quantity
\[ Q^n(\bs{\beta}):=
Q(\bs{Z}^{(n)}, \bs{\beta}, \alpha):=\sum_{i=1}^{n}r_i^2\mathds{1}\left( \frac{|r_i-m(\bs{Z}^{(n)},\bs{\beta})|}{\sigma(\bs{Z}^{(n)},\bs{\beta})}\leq \alpha\right),
\]
where $r_i=y_i-\bs{w}'_i\bs{\beta}$ and $m(\bs{Z}^{(n)},\bs{\beta})$ and $ \sigma(\bs{Z}^{(n)},\bs{\beta})$ are the median and MAD of $\{r_i\}$, respectively.
 We like to minimize $Q(\bs{Z}^{(n)}, \bs{\beta}, \alpha)$ over $\bs{\beta} \in \R^p$ (within a bounded ball centered at $\bs{\beta}_{lst}$ is sufficient) to obtain the least sum of squares of trimmed (LST) residuals estimator,
$$ \widehat{\bs{\beta}}^n_{lst}:=
\widehat{\bs{\beta}}_{lst}(\mb{Z}^{(n)}, \alpha)=\arg\min_{\bs{\beta}\in \R^p}Q(\bs{Z}^{(n)}, \bs{\beta}, \alpha).
$$
Based on Lemma 2.2,
it is readily seen that  $Q^n(\bs{\beta})$ is piece-wise twice continuously differentiable and convex in $\bs{\beta}\in \R^p$. By theorem 2.1, the solution is the least squares over some region $R_{\bs{\beta}^{k_0}}$, $1\leq k_0\leq L$.
\vs

\subsection{A procedure based Theorem 2.1} 

In light of Theorem 2.1, if one discovers all $R_{\bs{\beta}^k}$s for $1\leq k\leq L$, then one can get the exact result. But in practice and in some cases, this might be not affordable. However, one can simply search
as many $R_{\bs{\beta}^k}$s as possible to get a good approximation of the estimate $\widehat{\bs{\beta}}^{n}_{lst}$.
\vs
To identify $R_{\bs{\beta}^k}$ is equivalent to  identifying $i_1,\cdots, i_{K}$ so that $D_{i_1}< D_{i_2}<\cdots, D_{i_{K}}$ in light to (\ref{region.eqn}), where $K=|I(\bs{\beta}^k)|$. The latter is equivalent to finding a $\bs{\beta} \in R_{\bs{\beta}^k}$, then one gets the desired  $i_1,\cdots, i_{K}$.
 To find the desired $\bs{\beta}$, one way is to find a $\overline{\bs{\beta}}$ on the common boundary of $R_{\bs{\beta}^k}$ and $R_{\bs{\beta}^l}$ so that there are $i\neq j$, $D_i=D_j$ for some $0\leq l\neq k\leq L$ and $1\leq i, j \leq n$. Small perturbation of the coordinates of the $\overline{\bs{\beta}}=(\beta_1,\cdots, \beta_p)'$ leads to more than one $\bs{\beta}$s ($\bs{\beta}=(\beta_1,\cdots, \beta_j\pm \delta, \cdots, \beta_p)$ (for some $1\leq j\leq p$ and $\delta>0$) that  belong to $R_{\bs{\beta}^k}$ or $R_{\bs{\beta}^l}$.
 \vs
Now we address the way to find out $\overline{\bs{\beta}}$. In light of  (\ref{region.eqn}), there are $i\neq j$, $D_i=D_j$ for some $0\leq l\neq k\leq L$ and $1\leq i, j \leq n$. $D_i=D_j$ implies that (i) $r_i=r_j$ or (ii) $(r_i+r_j)/2=m_n(\bs{\beta})$. Both equalities could lead to some $\overline{\bs{\beta}}$s, but the first one $r_i=r_j$ is more convenient.
\vs
We now focus the first one which amounts to $y_i-y_j=(\bs{w}_i-\bs{w}_j)'\bs{\beta}=(\bs{x}_i-\bs{x}_j)'(\beta_2, \cdots, \beta_p)'$, where $\bs{w}'=(1,\bs{x}')$, $\bs{\beta}=(\beta_1,\cdots, \beta_p)$.  Assume that $\bs{x}_i\neq \bs{x}_j$ for $i\neq j$, if $y_i=y_j$, then, $\bs{\beta}=(\beta_1, \bs{0}'_{p-1})'$ is one of solutions,  otherwise,  from this equation, we see that (i) $\beta_1$ could be any number in $\R^1$, (ii) the equation defines a $(p-1)$-dimensional hyperplane. Consequently, all $\bs{\beta}=(\beta_1, 0,\cdots, 0, \frac{y_i-y_j}{x_{ik}-x_{jk}}, 0, \cdots, 0)\in \R^p$ are solutions, where $\beta_1\in \R^1$ and $x_{ik}\neq x_{jk}$, $1\leq k\leq (p-1)$. Simple choices for $\beta_1$ could be $0$ and $1$ or any constant.
From here we obtain at least two $\bs{\beta}$s that lie on the common boundary.
\vs
With the small perturbation ($\pm \delta$) to its ith coordinate of the $\bs{\beta}$s we could obtain $4p$ new $\bs{\beta}$s. For each such $\bs{\beta}$, we first obtain $i_1,\cdots, i_K$ with $K=|I(\bs{\beta})|$ and then check if the strictly inequalities in (\ref{region.eqn}) hold. \vs
 If they do not hold, then move to next $\bs{\beta}$. Otherwise, check if the $K$ indices already appear before,
 if it is, then do nothing, else update the data structure that stores indices, 
 and obtain the least square solution $\bs{\beta}_{ls}$-new based on the sub-data set with the $K$ subscripts ($I(\bs{\beta}$)) and the sum of squared residuals. If the latter is smaller than SS-min, then set it to be the SS-min and update $\widehat{\bs{\beta}}^n_{lst}$ with $\bs{\beta}_{ls}$-new. Increase $T_{ls}$, the counter for the number of LS calculation, by one. Move to next $\bs{\beta}$ until exhausting all $4p$ $\bs{\beta}$. Then repeat the entire process with a new pair $(i, j)$.
Summarizing discussions so far, we have
\vs
\noin
\tb{AA1-- pseudocode for LST based on Theorem 2.1}
\vs
\noin
\tb{Input}: A data set $\bs{Z}^{(n)}=\{(\bs{x}'_i, y_i)', i=1,2,\cdots, n\}$, a fixed $\alpha$. Assume that $\bs{x}_i\neq \bs{x}_j$ if $i\neq j$.\vs
\bi
\item[\tb{(1)}] Sample two indices $i$ and $j$ from  $\{1,\cdots, n\}$, 
assume that $x_{ik}\neq x_{jk}$, $1\leq k\leq (p-1)$ (i.e. the $k$th coordinates of $\bs{x}_i$ and $\bs{x}_j$ do not equal).  Consider
$$\bs{\beta}^0=(0, 0,\cdots, 0, \frac{y_i-y_j}{x_{ik}-x_{jk}}, 0,\cdots, 0)', \bs{\beta}^1=(1,0,\cdots, 0, \frac{y_i-y_j}{x_{ik}-x_{jk}}, 0,\cdots, 0)'\mbox{~in~}\R^p$$
Both have the same $(k+1)$th coordinate, $(y_i-y_j)/(x_{ik}-x_{jk})$.
\item[\tb{(2)}] Write $\bs{\beta}^j(l, \pm\delta)$  for the perturbed $\bs{\beta}^j$ with its $l$th coordinate adding or subtracting a $\delta>0$. Define a set
$$ S_{\bs{\beta}}=\cup_{l=1}^p\{\bs{\beta}^0(l, \pm \delta)\}\cup_{l=1}^p\{\bs{\beta}^1(l, \pm\delta)\}. $$

\item[\tb{(3)}]
For each $\bs{\beta}$ of $4p$ $\bs{\beta}$s is the set $S_{\bs{\beta}}$,
\bi
\item[\tb{(a)}] obtain $i_1,\cdots, i_K$ with $K=|I(\bs{\beta})|$ and check to see if the strictly inequalities in (\ref{region.eqn}) hold.
\bi 
\item[\tb{(a1)}] If not, move to the next $\bs{\beta}$; else\vs
\item[\tb{(a2)}] check if the $K$ indices already appear in a structure $S_{ind}$ 
\vs
\bi
\item[\tb{(i)}] if yes, then move to the next $\bs{\beta}$; else\vs
\item[\tb{(ii)}] update $S_{ind}$ by storing the $K$ indices in the structure $S_{ind}$ 
 and calculated LS estimate $\bs{\beta}_{ls}$-new based on the sub-data set with index in $I(\bs{\beta})$ and obtain the sum of $|I(\bs{\beta})|$ squared residuals, SS($\bs{\beta}_{ls}$-new). \vs
\item[\tb{(iii)}] Update $SS_{min}$ if it is greater than  SS($\bs{\beta}_{ls}$-new) and update $\widehat{\bs{\beta}}^n_{lst}$ with $\bs{\beta}_{ls}$-new. Update 
 the counter for the total number $T_{ls}$ of LS calculations, if the latter is less than $N$, then continue the loop (go to \tb{(3)}), else stop. 
\ei
\ei
\vs
\item[\tb{(b)}] If $T_{ls}<N$, then go to \tb{(1)}, else break the loop.
\ei
\ei
\vs
\noin
\tb{Output}: $\widehat{\bs{\beta}}^n_{lst}$
\vs
\noin
\tb{Remarks 6.1}~~ see the Appendix. \hfill\pend

\subsection{A  subsampling procedure}

Subsampling procedures are prevailing in practice for most robust regression (also location) estimators (see RL87,
 Hawkins 1994, Hawkins and Olive (1999), Rousseeuw and Struyf (1998), V\'{i}\v{s}ek (2001), RVD(1999, 2006), Zuo (2018, 2021c), among others).
 \vs
The basic idea is straightforward:
(1) draw a sub-sample of size $m$ from data set $\mb{Z}^{(n)}=\{(\bs{x'}_i, y_i)' \in \R^{p}, \bs{x}_i\in \R^{p-1}, i=1,2,\cdots, n\}$. (2) compute an estimate based on the sub-sample and obtain the objective function value.
(3) if the objective function value can be further improved (reduced), then go to (1), otherwise, stop and output the final step estimate.
\vs
Natural questions for the above procedure include (1) how to guarantee the convergence of the procedure and the final answer is the global minimum? (2) what is the exact size $m$ and what is the relationship with $n$ and dimension $p$? To better address these matters, we first propose the corresponding procedure for our LST.
\vs
\noindent
\tb{AA2} \tb{pseudocode for a sub-sampling procedure for LST}
\vs
\noin
\tb{Input}: A data set $\mb{Z}^{(n)}=\{\mb{Z}_1,\cdots, \mb{Z}_n \}=\{(\bs{x'}_i, y_i)', i=1,2,\cdots, n\}\in \R^{p}$ (assume that $p \geq 2$)  and an $\alpha\geq 1$ (default is one).
  \vs
\begin{enumerate}
\item[(a)]\tb{Initialization}:  N=$\min\{{n \choose p}, 300(p-1)\}$, R=0, $Q_{old}=10^8$, $\bs{\beta}_{old}=\mb{0}$ (or a LS (or LTS) estimate).
\item[(b)] \tb{Iteration}: while $(R\leq N)$
 \bi
\item[] keep sampling $p$ indices $\{i_1, \cdots, i_p\}$  from $\{1, 2, \cdots, n\}$ (without replacement) until $M'_{\bs{x}}:=(\bs{w}_{i_1}, \cdots, \bs{w}_{i_p})$ being invertible. Let $\bs{\beta}_{new}=(M_{\bs{x}})^{-1}(y_{i_1},\cdots, y_{i_p})'$.
   \bi
    \item[(1)] Calculate $I(\bs{\beta}_{new})$ (based on (\ref{I-beta.eqn})) and $Q_{new}:=Q^n(\bs{\beta}_{new})$ (based on (\ref{objective.eqn})).
    \item[(2)]
        \bi \item If $Q_{new}< Q_{old}$, then $Q_{old}=Q_{new}$, $\bs{\beta}_{old}=\bs{\beta}_{new}$. Get an LS estimator $\bs{\beta}_{ls}$ based on the data points of $\mb{Z}^{(n)}$ with subscripts from $I(\bs{\beta}_{new})$. Go to (1) with $\bs{\beta}_{new}=\bs{\beta}_{ls}$.
        \item Else if $Q_{new}=Q_{old}$ break\\
                ~~~~~~~ \hspace*{8mm}else R=R+1, go to (b)
        \ei
    \ei
\ei
\end{enumerate}
\tb{Output}: $\bs{\beta}_{new}$.  \hfill \pend
\vs
\noindent
\tb{Remarks 6.2} ~~ see the Appendix. \hfill\pend
\vs

\section{Examples and comparison}

This section investigates the performance of AAs and compares it with that of the benchmark LTS. First, we like to give some guidance for selection among the two AAs.\vs
\noin

\tb{Example 7.1} \tb{Performance of the two AAs} ~ There are two AAs and which of them should be recommended for users?
This example tries to achieve this by examining the speed and accuracy of the two AAs. \vs
We generate $1000$ samples $\mb{Z}^{(n)}=\{(\bs{x}'_i, y_i), i=1,\cdots, n, \bs{x}_i \in \R^{p-1}\}$ from the  standard Gaussian  distribution
for various sample size $n$ and dimension $p$. For the speed, we calculate the \emph{total time} consumed for all $1000$ samples (dividing it by $1000$, one gets the average time consumed per sample) by different AAs. For accuracy (or variance, or efficiency), we will compute their empirical mean squared error (EMSE).\vs

For a general estimator $\mb{T}$, if it 
 is  regression equivariant, then we can assume (w.l.o.g.) that the true parameter $\bs{\beta}_0=\mb{0}\in \R^p$.  We calculate
$\mbox{EMSE}:=\sum_{i=1}^R \|\mb{T}_i - \bs{\beta}_0\|^2/R$, the empirical mean squared error (EMSE) for $\mb{T}$, where
$R = 1000$, $\bs{\beta}_0 = (0, \cdots, 0)'\in \R^{p}$,
 and $\mb{T}_i$ is the realization of $\mb{T}$ obtained from the ith sample with size $n$ and dimension $p$. The EMSE and the total time consumed (in seconds)  by different AAs are listed in Table \ref{table-comp-time-AAs}.

\begin{table}[t!]
\centering
~~ Table entries (a, b) are: a:=empirical mean squared error, b:=total time consumed  
\bec
\begin{tabular}{c c c c  }
~n~ & ~p~& ~~~~AA1~~~~ & ~~~~AA2~~~~\\
\hline\\[0.ex]

 &3&(0.3499,~~566.49) &(0.5290,~~651.25) 
\\
50  &5&(0.5817,~~457.49) &(0.7645,~~861.75) 
\\  &10 &(0.5390,~~682.41) &(1.7177,~~1016.6) 
  \\[2ex]

 &3&(0.1755,~~573.07) &(0.3619,~~879.01) 
\\
100  &5&(0.2023,~~638.76) &(0.4528,~~1042.6) 
\\
&10 &(0.2576,~~702.02) &(0.7000,~~1071.5) 
  \\[2ex]

 &3&(0.0825,~~619.75) &(0.3025,~~1309.7) 
\\
200  &5&(0.1055,~~676.63) &(0.3501,~~1285.6) 
  \\&10 &(0.1283,~~698.14) &(0.4178,~~1310.2) 
  \\[1ex]
\hline
\end{tabular}
\enc
\caption{Total computation time  for all $1000$ samples (seconds) and empirical mean squared error (EMSE) of different AAs for various $n$s and $p$s.}
\label{table-comp-time-AAs}
\end{table}
 \vs

Inspecting Table \ref{table-comp-time-AAs} immediately reveals that (i) AA2 is not only the slowest but is most inaccurate (with  the largest EMSEs) in all cases considered. (ii) AA1 has both speed and accuracy advantages for all cases considered.
\vs
 Overall, we recommend AA1 for users. That does not exclude the potential of improvement of AA2 via the idea in Rousseeuw and Van Driessen (2006).\hfill \pend

 \vs
All R code for simulation and examples as well as figures in this article (downloadable via https://github.com/zuo-github/lst) were run on a desktop Intel(R)Core(TM)
21 i7-2600 CPU @ 3.40 GHz.\vs
The data points in the example above are perfect standard normal and hence are not practically realistic. In the following, we will investigate the performance of AA1 versus LTS for contaminated  standard normal data sets and for moderate as well as large $n$s and $p$s.
\vs
\noin
\vs
\tb{Example 7.2 ~Multiple regression with contaminated normal data sets}.
Now we consider data with contamination, which is typical for big data sets in ``big-data-era".  \vs
We consider the contaminated highly correlated normal data points scheme. We generate $1000$ samples $\mb{Z}_i=(\bs{x_i}',y_i)'$ with various $n$s  from the  normal distribution $\mc{N}(\bs{\mu}, \bs{\Sigma})$,
where $\bs{\mu}$ is a zero-vector in $\R^p$, and $\bs{\Sigma}$ is a $p$ by $p$ matrix with diagonal entries being $1$ and off-diagonal entries being $0.9$. Then $\varepsilon\%$ of them are contaminated by normal points
with $\bs{\mu}$ being the $p$-vector with all elements being $7$ except the last one being $-2$ and the covariance matrix being diagonal with diagonal being $0.1$. The results are listed in Table \ref{lts-vs-lst-alpha-30} .
\vs
\begin{table}[!h]
\centering
~~ Normal data sets, each with $\varepsilon \%$ contamination\\
~~ Table entries (a, b) are: a:=empirical mean squared error,  b:=total time consumed  
\bec
\begin{tabular}{c c c c c c }
~ & ~& ~~~~~~~~~~~~~$\varepsilon=5\%$ &&~~~~~~~~~~~~$\varepsilon=10\%$ &\\
~p~ & ~n~& ~~AA1~~  &~~ltsReg~~& AA1~~&ltsReg \\
\hline\\[0.ex]
 & 100& (0.2971,~~9.6581)& (0.3010,~~22.867)& (0.2843,~~494.01)& (0.2942,~~ 25.289)\\
5&200&(0.2503,~~26.045)& (0.2650,~~41.861)& (0.2517,~~26.629)&(0.2630,~~ 43.504)\\
& 300&(0.2396,~~54.100)& (0.2551,~~63.639)& (0.2366,~~54.885)&(0.2534,~~ 63.522)\\[2ex]
 & 400& (0.1335,~~1085.6)& (0.1394,~~181.18)& (0.1340,~~1056.2)&(0.1382,~~175.92)\\
10&500&(0.1280,~~1207.7)& (0.1321,~~222.81)& (0.1289,~~1178.5)&(0.1321,~~218.94)\\
& 600&(0.1247,~~1308.4)& (0.1285,~~152.47)& (0.1253,~~1273.6)&(0.1276,~~149.99)\\[2ex]
 & 700& (0.0815,~~2044.9)& (0.0885,~~549.61)& (0.0838,~~1994.0)&(0.0882,~~547.53)\\
20&800&(0.0776,~~2261.7)& (0.0837,~~620.63)& (0.0796,~~2177.0)&(0.0837,~~616.87)\\
& 900&(0.0748,~~2436.1)& (0.0804,~~541.20)& (0.0761,~~2353.7)& (0.0795,~~ 538.43)\\[2ex]
~ & ~& ~~~~~~~~~~~~~$\varepsilon=30\%$ &&~~~~~~~~~~~~$\varepsilon=40\%$ &\\
&300&(0.4347,~~53.248)& (1.9236,~~1635.1)& (0.4352,~~56.430)& (1.3517,~~1712.8)\\
40&400&(0.3362,~~100.04)& (1.2604,~~2401.5)& (0.3314,~~102.81)&(0.8995,~~2399.5)\\
&500&(0.2594,~~147.66)& (0.9514,~~ 2963.4)& (0.2873,~~146.67)&(0.6851,~~2787.7)\\[2ex]
&300&(0.5242,~~58.736)& (2.7826,~~2861.8)& (0.5700,~~59.903)&(1.9808,~~2896.3)\\
50&400&(0.4085,~~ 89.897) &(1.7562,~~3292.0)& (0.4539,~~108.88)&(1.2547,~~3925.5)\\
&500&(0.3107,~~145.84)& (1.2870,~~4510.5)& (0.3406,~~145.75)&(0.9086,~~4419.6)\\[0ex]
\hline
\end{tabular}
\enc
\caption{Total computation time  for all $1000$ samples (seconds) and empirical mean squared error (EMSE) of LST(AA1) versus LTS(ltsReg) for various $n$s, $p$s, and contaminations.}
\label{lts-vs-lst-alpha-30}
\end{table}

Inspecting the table reveals that (i) in terms of EMSE, AA1 is the overall winner (with the smallest EMSE in all cases considered), LTS has the largest EMSE in all the cases (this is not surprising since if exhausting all $L$ pieces in Theorem 2.1, one can get exact result from AA1);
 (ii) in terms of speed, LTS (or rather ltsReg) is the winner when $d=10$ or $20$. AA1 is the winner for all other $d$'s, except when $d=5$, $n=100$ and $\varepsilon=10\%$. For the latter case, AA1 can still be the faster if tuning $T_{ls}$ to be $1$, then one gets $(0.2986,10.396)$ for AA1 versus $(0.2948,23.133)$ for ltsReg (suffering a slight increase in EMSE).
   \hfill \pend
   \vs
LTS (or lstReg) demonstrates its well-known  speedy advantage, which is  partially due to its background computation via Fortran subroutine and the computation scheme proposed in RVD06. AA1 (a pure R programming procedure), on the other hand, has the potential to speed up via Rcpp or even via Fortran in one or more order of magnitude.
\vs
\noin
\tb{Remarks 7.1}
\vs
\tb{(I)} \tb{Parameters tuning}~ Two parameters in AA1 that can be tuned. 
The $T_{ls}$ is set to be $300$ for better EMSE (as in the $d=5$, $n=100$, and $\varepsilon=10\%$ case). If tuning it to be $1$, one gets a much faster AA1 (as in the cases $d=30$ $40$, and $d=5$, except when $n=100$, and $\varepsilon=10\%$).
 For the $\alpha$ in the definition of LST, it is set to be $1$ (default value) in Table $1$, it is set to be $3$ as in Table $2$ when there are contaminations (or outliers). Note that theoretically speaking, both LST and LTS can resist $50\%$ contamination without breakdown. So $40\%$ contamination rate in Table $2$ is relevant which is also employed in RVD06.  \vs

\tb{(II)} LTS estimate is obtained via R package ltsReg, $h$ is the  default value $\lfloor (n+p+1)/2\rfloor$, one might tune this $h$ to get better performance from LTS. But this will decrease LTS's finite sample breakdown value. This is not the case for LST with the $\alpha$ (see Theorem 3.1).
\hfill \pend
\vs
Up to this point, we have dealt with synthetic data sets. Next we investigate the performance of LST and LTS with respect to real data sets in high dimensions.
\vs
\tb{Example 7.3}~~ \tb{Textbook size real data sets} We first look at real data sets with relatively small sample size $n$  and moderate dimension $p$. For a description of data sets, see RL87, all are studied there.
Since all methods depend on randomness, So we run the computation with replication number $R=1000$ times to alleviate the randomness (in light of the LLN),
 we then calculate the \emph{total} time consumed (in seconds) by different methods for all replications, and the EMSE (with true $\bs{\beta}_0$ being replaced by the sample mean of $1000$ $\widehat{\bs{\beta}}$s), which is the sample variance of all $\widehat{\bs{\beta}}$s up to a factor $1000/999$.
The results are reported in Table \ref{lts-vs-lst-alpha-3-real-data-0}, where the parameters $\alpha$ and $T_{ls}$ in AA1 are tuned.

\begin{table}[!h]
\centering
Table entries (a, b) are: a:=empirical variance of $\widehat{\bs{\beta}}$s,  b:=total time consumed
\bec
\begin{tabular}{c c c c }
data set &~(n,~p)~& ~~~~AA1~~ ~~ &~~~~ltsReg~~~~\\
\hline\\[0.ex]
salinity &(28, ~4) &(184.60,~~2.3700)   &(2220.1,~~8.9105)\\[2ex]
aircraft &(23,~5) &(60.194,~~32.041) &(178.56,~~10.037)\\[2ex]
wood  &(20,~6) &(1.4903,~~5.0559)  &(2.0821,~~10.448)\\[2ex]
coleman &(20,~6) &(530.22,~~17.386) &(1588.7,~~10.376)\\[2ex]
\hline
\end{tabular}
\enc
\caption{Total time consumed (in seconds) and sample variance in $1000$ replications by LTS (ltsReg) and LST (AA1) for various real data sets.}
\label{lts-vs-lst-alpha-3-real-data-0}
\end{table}

Inspecting the Table reveals that (i) in terms of the empirical mean squared error (or rather empirical variance), AA1 is the over-all winner for all cases considered and LTS has the largest sample variance. (ii) in terms of computation speed, AA1 and ltsReg have equal shares, both win two cases among the four. The latter is faster for data sets: aircraft and coleman. \hfill \pend
\vs The limitation of this example is that the data sets are still relatively small and not in very high dimensions. We examine a high dimension and large sample dataset next.
\vs
\tb{Example 7.4}~~ \tb{A large real data set}~  Boston housing is a famous data set (Harrison, D. and Rubinfeld, D.L. (1987)) and studied by many authors with different emphasizes (transformation, quantile, nonparametric regression, etc.) in the literature. For a more detailed description of the data set, see http://lib.stat.cmu.edu/datasets/.\vs
 The analysis
reported here did not include any of the previous results, but consisted of just a straight linear regression of the dependent variable (median price of a house) on the thirteen explanatory variables
as might be used in an initial exploratory analysis of a new data set.
 We have sample size $n=506$ and dimension $p=14$.
\vs
 Our scheme to evaluate the performance of LST and LTS is as follows: (i) we sample $m$ points (without replacement) ($m=506$, entire data set, or $m=200, 250, 300, 350$) from the entire data set,
and compute the $\widehat{\bs{\beta}}$s with different methods, we do this RepN times, where replication number RepN varies with respect to different $m$s.
(ii) we calculate the total time consumed (in seconds) by different methods for all replications, and the EMSE (with true $\bs{\beta}_0$ being replaced by the sample mean of RepN $\widehat{\bs{\beta}}$s from (i)), which is the sample variance of all $\widehat{\bs{\beta}}$s up to a factor $RepN/(RepN-1)$.
The results are reported in Table \ref{lts-vs-lst-alpha-3-real-data}.\vs

\begin{table}[!h]
\centering
Table entries (a, b) are: a:=empirical variance of $\widehat{\bs{\beta}}$s,  b:=total time consumed
\bec
\begin{tabular}{c c c c c }
p&m & RepN & ~~AA1~~  & ~~ltsReg~~\\[1ex]
\hline\\[.0ex]
&200 & $10^4$& (779.61,~~289.91) & (551.39,~~513.46)\\
&250 & $10^4$& (762.98,~~439.71) & (520.48,~~674.29)\\
14&300 & $10^4$& (751.31,~~633.39)& (519.25,~~785.87)\\
&350 & $10^4$& (764.67,~~818.20) &  (515.23,~~901.98)\\
&506 & $10^3$& (173.45,~~149.59) &  (480.92,~~119.55)\\
\hline
\end{tabular}
\enc
\caption{Total time consumed (in seconds) and sample variance in RepN replications by LTS (ltsReg) and LST (AA1) for real data sets with various sample size $m$'s and $p=14$.}
\label{lts-vs-lst-alpha-3-real-data}
\end{table}

Inspecting the Table reveals that (i) ltsReg has the smallest sample variance in all cases considered but with a price of being the slowest (with the exception when $m=506$);
(ii) AA1 is faster than ltsReg but with a price of having  slightly larger sample variances (with the exception when $m=506$). \hfill \pend
\vs
\section{Final discussions}

\tb{The difference between LTS and LST}~~
The least sum of squares of trimmed (LST) residuals  estimator, which is proven to have the best $50\%$ asymptotic breakdown point, is another robust alternative to the classical least sum of squares (LS) of residuals estimator.  The latter keeps all squared residuals whereas the former trims some residuals then squares the left. Trimming is also utilized  in the prevailing least sum of trimmed squares (LTS) of the residuals  estimator. However, the two trimming schemes are quite different,  the one used in LTS is a one-sided trimming (only large squared residuals are trimmed, of course, it also might be regarded as a two-sided trimming with respect to the un-squared residuals) whereas the one utilized in LST is a depth-based trimming (see Zuo (2006) and Wu and Zuo (2009) for more discussions on  trimming schemes) which can trim both ends of un-squared residuals and trim not a fixed number of residuals. \vs Besides the trimming scheme difference, there is another  difference between LTS and LST, that is, the order of trimming and squaring. In LTS, squaring is first, followed by trimming whereas in LST, the order is reversed. All the difference leads to an unexpected performance difference in LTS and LST as demonstrated in the last section.
\vs
\noin
\tb{The status of the art}~~The idea of trimming residuals and then doing regression has appeared in the literature for quite some time. The trimming idea was first introduced in location setting  but later extended to regression, see, Huber (1973), Bickel (1975), Ruppert and Carroll (1980), Welsh(1987), and RL87, among others. However, trimming residuals based on depth or outlyingness employed in this article is novel and never utilized before. A more recent study on the topic is given in Johansen and Nielsen (2013), where the authors used an iterated one-step
approximation to the Huber-skip estimator to detect outliers in regression, and theoretical justification for the approximation is provided. Their Huber-skip estimator defined on page 56 is closely related to our LST,
but has two essential differences (i) their estimator more resembles the least winsorized squares regression (see page 135 of RL87), (ii) residuals in their estimator are not centered by the median of residuals.
\vs
\noindent
\tb{Fairness of performance criteria}~~ For comparison of the performance of LST and LTS, we have focused on the variance (accuracy, efficiency, or EMSE) and the computation speed of the algorithms for the estimators.
The asymptotic efficiency (AE) of LTS has been reported to be just $7\%$  in Stromgberg, et al (2000) or $8\%$ in MMY06 (page 132), the AE of LST is yet to be discovered, which however is expected to be better than $8\%$. This assentation is verified and supported by the experimental results in the last section (Tables 2, and 3 indicate that the LST is much more efficient than the LTS). Furthermore, it was also supported by the results in Wu and Zuo (2009) for various trimming schemes in the case of $p=1$.
\vs
The computation speed comparison of LTS versus LST in the last section is somewhat not based on a fair ground. It is essentially a speed comparison of pure R verse R plus Fortran since the  Fortran subroutine (rfltsreg) is called in ltsReg.
Even with that, ltsReg does not have an overwhelming advantage on speed over AA1. For the latter, however, there is still  room for improvement by utilizing Fortran or even better Rcpp to speed up by at least one order of magnitude.\vs

\noindent
\tb{Connection with notion of depth in regression and regression medians} According to Zuo (2021a), both LTS and LST could be regarded as a deepest estimator (a regression median) with respect to the corresponding
objective function type of regression depth (see Section 2.3.1 of Zuo (2021a)). \vs

\noindent

\noin
\tb{Parameters tuning and finite sample breakdown point} There are two parameters $h$ in LTS and $\alpha$ in LST which can be tuned in the program for computation. Their values have a connection with the finite sample breakdown point. For example, when $h$ takes its default value $\lfloor(n+p+1)/2\rfloor$, then the FSBP of LTS is $(n-h+1)/n$ which will decrease from the best FSBP result $(\lfloor(n-p)/2\rfloor+1)/n$ (see pages 125, 132 of RL87) when $h$ increases. For the parameter $\alpha$ in LST, as long as $\alpha \geq 1$ then the high FSBP in theorem 3.1 remains valid. This is due to the difference in the trimming schemes (see Wu and Zuo (2019)). \vs
\noindent
\tb{Open and future problems}~~ 
 By simply switching the order of trimming and squaring and adopting a depth based trimming scheme, LTS and LST can have such different performance. One naturally wonders what if one does the same thing with respect to the famous LMS introduced also by Rouseeuw (1984) (i.e. the least square of the median (LSM) of residuals  estimator). It turns out, this is not a good idea since there is a universal solution,  it is  $\widehat{\bs{\beta}}=(\mbox{Med}\{y_i\}, 0,\cdots, 0)\in R^p$.\vs One interesting problem that remains is to investigate the least sum of squares of trimmed residuals with yet another trimming scheme such as the winsorized version given in Wu and Zuo (2019), that is, replacing the residuals beyond the cut-off values at the two ends with just the cutoff values or even a more generalized weighted (trimming) scheme which includes the hard $0$ and $1$ trimming scheme.
 Other challenging open topics that deserve to be pursued independently elsewhere include (i) providing a finite sample estimation error analysis (non-asymptotic analysis) (ii) regularized regression based on the LST to handle variable selection and model interpretation issues when dimension $p$ is much larger than sample size $n$.
\vs
\vs
\noin
{\textbf{\Large Acknowledgments}}
\vs
Authors thank Denis Selyuzhitsky,  Nadav Langberg, and Profs. Wei Shao and Yimin Xiao for their
 insightful comments and stimulating discussions which significantly improved the manuscript.

\vs
\noin
{\textbf{\Large Declarations}}
\vs
\noin
\tb{Funding}

Authors declare that there is no funding received for this study.
\vs
\noin
\tb{Conflicts of interests/Competing interests}

Authors declare that there is no conflict of interests/Competing interests.

\vs
\bigskip
\hspace*{20mm} {\large\bf SUPPLEMENTARY MATERIAL}

\begin{description}

\item [R code downloadable at https://github.com/left-github-4-codes/lst]

\noindent
\item [Appendix: main proofs and remarks]
\noindent
\vs
\tb{Proof of Lemma 2.2}
\vs
\tb{(i)} For $\bs{\eta} \in R_{\bs{\beta}^k}$, we have $I(\bs{\eta})=I(\bs{\beta}^k)$. Let $J=|I(\bs{\beta}^k)|$, then $D_{i_{j+1}}(\bs{\eta})>D_{i_{j}}(\bs{\eta})$ for $1\leq j\leq  (J-1)$. Let $\gamma :=\min_{1\leq j\leq  (J-1)}|D_{i_{j+1}}(\bs{\eta})-D_{i_j}(\bs{\eta})|\sigma_n(\bs{\eta})$, then by (\ref{region.eqn}) we have $\gamma>0$. \vs 
Due to the continuity of residuals in $\bs{\beta}$, we can choose a small radius $\delta$ such that for any $\bs{\beta}\in B(\bs{\eta}, \delta)$,
$|r_i(\bs{\eta})-r_i(\bs{\beta})|<\gamma/4$ for any  $i$.  
A straightforward derivation one gets $|m_n(\bs{\eta})-m_n(\bs{\beta})|\leq \gamma/4$. 
In light of these two inequalities and the definition of $\gamma$, one obtains
\begin{align*}
|r_{i_{j+1}}(\bs{\beta})-m_n(\bs{\beta})|&\geq \big|r_{i_{j+1}}(\bs{\eta})-\gamma/4-[m_n(\bs{\eta})+\gamma/4]\big|\\[1ex]
&=\big|r_{i_{j+1}}(\bs{\eta})-m_n(\bs{\eta})-\gamma/2\big|\\[1ex]
&\geq |r_{i_{j}}(\bs{\eta})-m_n(\bs{\eta})|+\gamma/2,
\end{align*}
for any $\bs{\beta}\in B(\bs{\eta}, \delta)$ and any $1\leq j\leq  (J-1)$, and
\begin{align*}
|r_{i_{j}}(\bs{\beta})-m_n(\bs{\beta})|&\leq |r_{i_{j}}(\bs{\eta})+\gamma/4-[m_n(\bs{\eta})-\gamma/4]|\\
&=|r_{i_{j}}(\bs{\eta})-m_n(\bs{\eta})+\gamma/2|\\
&\leq |r_{i_{j}}(\bs{\eta})-m_n(\bs{\eta})+ \gamma/2
\end{align*}
The last two displays imply that $D_{i_{j+1}}(\bs{\beta})> D_{i_{j}}(\bs{\beta})$ for any $1\leq j\leq  (J-1)$. That is, for any $\bs{\beta}\in B(\bs{\eta}, \delta)$, $\bs{\beta}\in R_{\bs{\beta}^k}$. Consequently, $Q^n(\bs{\beta})=\sum_{i\in I(\bs{\beta}^k)}r^2_i$.\vs
\tb{(ii)} The openness of $R_{\bs{\beta}^k}$ follows from the proof (i) above straightforwardly.
\vs
\tb{(iii)} For any $\bs{\beta}\in \R^p$, (i) either $\bs{\beta}\in R_{\bs{\beta}^k}$ for some $0\leq k\leq L$ and $Q^n(\bs{\beta})=\sum_{i\in I(\bs{\beta})}r^2_i$, or (ii) $\bs{\beta}$ lies on the common boundary of $R_{\bs{\beta}^s}$ and $R_{\bs{\beta}^t}$ for some $1\leq s\neq t\leq L$ such that
there are $i\neq j$ $D_i(\bs{\beta})=D_j(\bs{\beta})$, and $D_i(\bs{\eta})>D_j(\bs{\eta})$ if $\bs{\eta} \in R_{ \bs{\beta}^s}$ and $D_i(\bs{\eta})<D_j(\bs{\eta})$ if $\bs{\eta} \in R_{ \bs{\beta}^t}$, and $Q^n(\bs{\beta})=\sum_{i\in I(\bs{\beta})}r^2_i$ for $\bs{\beta}\in \overline{R}_{\bs{\beta}^s}\cap \overline{R}_{\bs{\beta}^t}$. \vs  The continuity of $Q^n(\bs{\beta})$ over $R_{\bs{\beta}^k}$ is obvious. We show that is true at any $\bs{\eta}\in \overline{R}_{\bs{\beta}^s}\cap \overline{R}_{\bs{\beta}^t}$. 
Let $\{\bs{\beta}_j\}$ be a sequence approaching to $\bs{\eta}$, where $\bs{\beta}_j$ could be in $\overline{S}_{\bs{\beta}^{s}}$ or in $\overline{S}_{\bs{\beta}^{t}}$. We show that $Q^n(\bs{\beta}_j)$ approaches to $Q^n(\bs{\eta})$. 
Note that $Q^n(\bs{\eta})=\sum_{i\in I(\bs{\eta})}r^2_i$ for $\bs{\eta}\in \overline{R}_{\bs{\beta}^s}\cup \overline{R}_{\bs{\beta}^t}$. Partition $\{\bs{\beta}_j\}$ into $\{\bs{\beta}_{j_s}\}$ and $\{\bs{\beta}_{j_t}\}$, and all members of the former belong to $\overline {R}_{\bs{\beta}^s}$ where the latter are all within $\overline{R}_{\bs{\beta}^{t}}$.
By continuity of the sum of squared residuals in $\bs{\beta}$, both $Q^n(\bs{\beta}_{j_s}))$ and $O^n(\bs{\beta}_{j_t}))$ approach to $Q^n(\bs{\eta})$ since both $\{\bs{\beta}_{j_s}\}$ and $\{\bs{\beta}_{j_t}\}$ approach $\eta$ as $\min \{j_s, j_t\} \to \infty$.
\vs
\tb{(iv)} Over each $R_{\bs{\beta}^k}$, $1\leq k\leq L$, $Q^n(\bs{\beta})=\sum_{i\in I(\bs{\beta})}r^2_i$ which is clearly twice differentiable and convex since
\begin{align*}
\frac{\partial}{\partial\bs{\beta}} Q^n(\bs{\beta}) &= -2\sum_{i=1}^nr_i \mathds{1}_i\bs{w}_i=-2\bs{R}'\bs{D}\bs{W}_n', \\[1ex]
\frac{\partial^2}{\partial\bs{\beta}^2} O^n(\bs{\beta})&=2\bs{W}_n\bs{D}\bs{W}_n',
\end{align*}
where $\bs{R}=(r_1, r_2, \cdots, r_n)'$, $\bs{D}=\mbox{diag}(\mathds{1}_i)$, and $\bs{W}_n=(\bs{w}_1, \bs{w}_2, \cdots, \bs{w}_n)'$. Strict convexity follows from the positive definite of Hessian matrix: ${2}\bs{W}_n\bs{D}\bs{W}_n'$.
\hfill \pend

\vs
\noin
\tb{Proof of Theorem 2.1}
\vs
\noindent
{(i)}
Over each ${S}_{\bs{\beta}^k}$, $Q^n(\bs{\beta})$ is twice differentiable and strictly convex in light of given condition,
 hence it has a unique minimizer. Since there are only finitely many ${R}_{\bs{\beta}^k}$,  the assertion follows if we can prove that the minimum does not reach at a boundary point of some ${R}_{\bs{\beta}^k}$.
\vs
 Assume it is otherwise. That is, $Q^n(\bs{\beta})$ reaches its minimum at point $\bs{\beta}_1$ which is a boundary point of $R_{\bs{\beta}^k}$ for some $k$. Assume that over $R_{\bs{\beta}^k}$, $Q^n(\bs{\beta})$ attains its minimum value at the unique point $\bs{\beta}_2$. Then, $Q^n(\bs{\beta}_1)\leq Q^n(\bs{\beta}_2)$, If equality holds then
we already have the desired result, otherwise, there is a point $\bs{\beta}_3$ in the small neighborhood of $\bs{\beta}_1$ so that $Q^n(\bs{\beta}_3)\leq Q^n(\bs{\beta}_1)+(Q^n(\bs{\beta}_2)-Q^n(\bs{\beta}_1))/2< Q^n(\bs{\beta}_2)$. A contradiction is obtained.
\vs
{(ii)} It is seen from \tb{(i)} that $Q^n(\bs{\beta})$ is twice continuously differentiable, hence its first derivative evaluated at the global minimum must be zero. By \tb{(i)}, we have 
(\ref{estimation.eqn}). \vs

{(iii)} This part directly follows from \tb{(ii)} and the invertibility of $\bs{M}_n$ that follows from the full rank of $\bs{X}_n$. 
\hfill
\pend
\vs
\vs
\noindent
\tb{Proof of Theorem 2.2}\vs
For the given  $\bs{Z}^{(n)}$ and $\alpha$, 
 write $M=Q(\bs{Z}^{(n)}, \bs{0}, \alpha)=\sum_{i\in I(\bs{0})} y^2_i $.
For a given $\bs{\beta}\in\R^p$, assume that $H_{\bs{\beta}}$ is the hyperplane determined by $y=\bs{w}'\bs{\beta}$ and let $H_h$ being the horizontal hyperplane (i.e. $y=0$, the $\bs{w}$-space).
Partition the space of $\bs{\beta}$s into two parts:$S_1$ and $S_2$, with $S_1$ containing all $\bs{\beta}$s such that $H_{\bs{\beta}}$ and $H_h$ are parallel and $S_2$ consisting of the rest of  $\bs{\beta}$s so that $H_{\bs{\beta}}$ and $H_h$ are not parallel.
\vs
If one can show that there are minimizers of $Q(\bs{Z}^{(n)}, \bs{\beta}, \alpha)$ over $S_i$ $i=1,2$ respectively, then one can have an overall minimizer. Over $S_1$,
the minimizer is $\widehat{\bs{\beta}}=(\overline{y}, \mb{0}'_{(p-1)\times 1})'$ and the minimum value of $Q(\bs{Z}^{(n)}, \widehat{\bs{\beta}}, \alpha)$ is $M-\overline{y}^2$, where $\overline{y}$ is the average of $y_i$ over all $i\in I(\bs{0})$.
\vs

Over $S_2$, denote by $l_{\bs{\beta}}$  the
intersection part of $H_{\bs{\beta}}$ with the horizontal hyperplane $H_h$ 
 (we call it a hyperline, though it is $p-1$-dimensional).
 Let $\theta_{\bs{\beta}}\in(-\pi/2, \pi/2)$ be the angle between the $H_{\bs{\beta}}$ and $H_h$ (and $\theta_{\bs{\beta}}\not =0$).
Consider two cases. 
\vs
\tb{Case I}. All $\bs{w}_i$ , $i\in I(\bs{\beta})$ on the hyperline $l_{\bs{\beta}}$. Then we have a vertical hyperplane that is perpendicular to the horizontal hyperplane $H_h$ $(y=0)$ and intersect $H_h$ at
$l_{\bs{\beta}}$, which contains, in light of lemma 2.1,
at least $\lfloor (n+1)/2\rfloor$ points of $\bs{Z}^{(n)}$.  But this contradicts the assumption just before the theorem. We only need to consider the other case.
\vs
\tb{Case II}. Otherwise, define
$$
\delta=\frac{1}2\inf\{\tau, \mbox{such that $N(l_{\bs{\beta}}, \tau)$ contains all $\bs{w}_i$ with $i\in I(\bs{\beta})$}
\},
$$
where $N(l_{\bs{\beta}}, \tau)$ is the set of points in $\bs{w}$-space such that each distance to the $l_{\bs{\beta}}$ is no greater than $\tau$. Clearly, $0<\delta<\infty$ (since $\delta=0$ has been covered in \tb{Case I} and $2\delta\leq \max_i\{\|\bs{w}_i\|\}<\infty$, where the first inequality follows from the fact that hypotenuse is always longer than any legs).
\bec
\vspace*{-15mm}
\begin{figure}[t]
\includegraphics
[scale=0.75]
{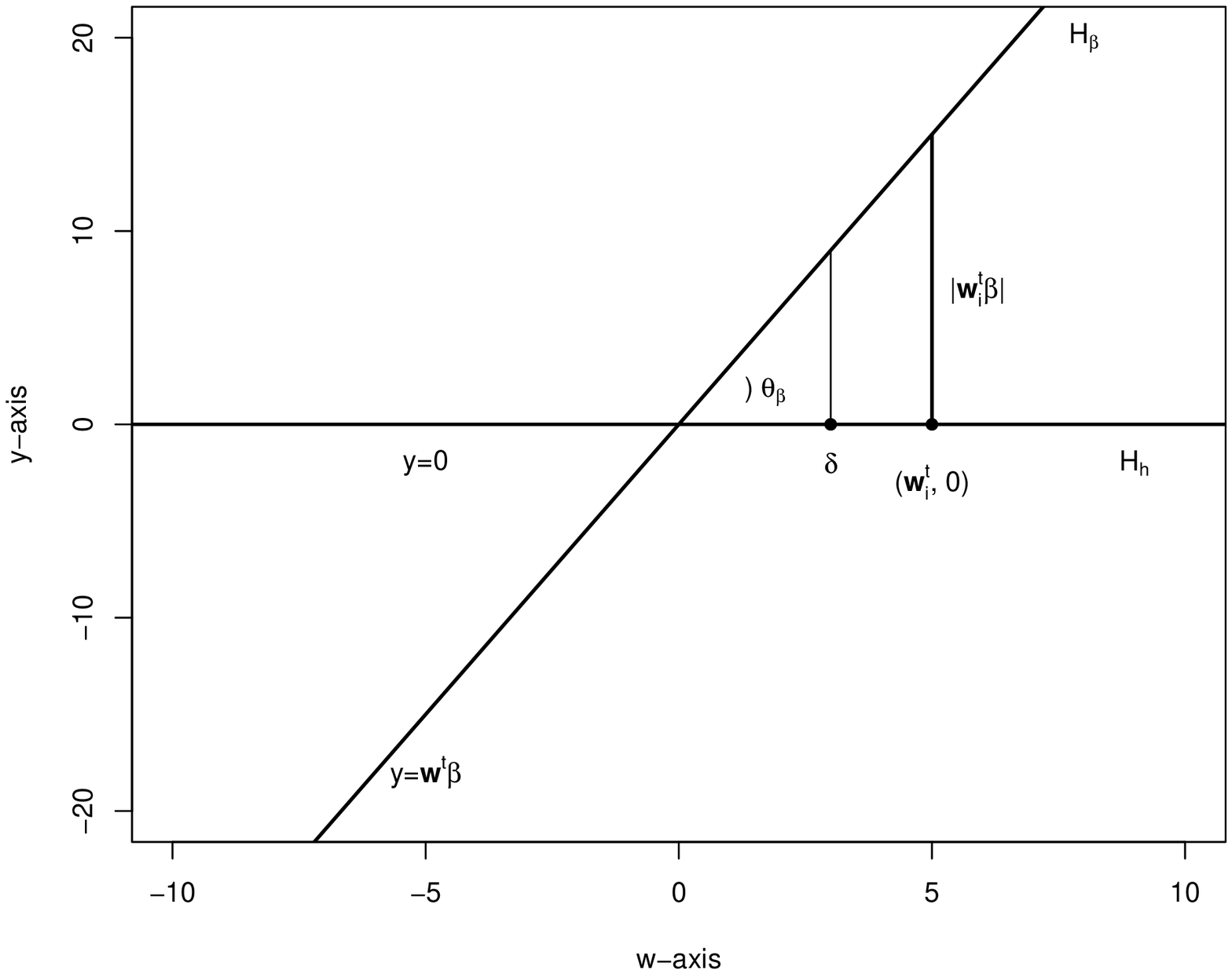}
    \caption{A two-dimensional vertical cross-section (that goes through points $(\bs{w}^t_i, 0)$ and $(\bs{w}^t_i, \bs{w}_i^t\bs{\beta})$) of a figure in $\R^p$ ($\bs{w}_i^t=\bs{w}'_i$).
    Hyperplanes $H_h$ and $H_{\bs{\beta}}$ intersect at hyperline $l_{\bs{\beta}}$ (which does not necessarily pass through $(\bs{0},0)$, here just for illustration).
     The vertical distance from point $(\bs{w}^t_i, 0)$ to the hyperplane $H_{\bs{\beta}}$, $|\bs{w}^t_i\bs{\beta}|$, is greater than $\delta|\tan(\theta_{\bs{\beta}})|$. }
  \label{fig-4-proof}
  \vspace*{-2mm}
\end{figure}
\enc
\vs

 We now show that when $\|\bs{\beta}\|> (1+\eta)\sqrt{M}/\delta$, where $\eta>1$ is a fixed number, then
\be
\sum_{i\in I(\bs{\beta})}r^2_i(\bs{\beta})>M=\sum_{i\in I(\mb{0})}r^2_i(\bs{0}).\label{proof.inequality}
\ee
That is, for the solution of minimization of (\ref{objective.eqn}), one only needs to search over the ball $\|\bs{\beta}\|\leq(1+\eta)\sqrt{M}/\delta$, a compact set.
 Note that $Q(\bs{Z}^{(n)}, \bs{\beta}, \alpha)$ is continuous in
$\bs{\beta}$ by Lemma 2.2.
Then the minimization problem certainly has a solution over the compact set.
\vs
The proof is complete if we can show (\ref{proof.inequality}) when $\|\bs{\beta}\|> (1+\eta)\sqrt{M}/\delta$. It is not difficult to see that there is at least one $i\in I(\bs{\beta})$ such that $\bs{w}_i \not \in N(l_{\bs{\beta}}, \delta)$ since otherwise it contradicts the definition of $\delta$ above.  Note that $\theta_{\bs{\beta}}$ is the angle between the normal vectors $(-\bs{\beta}', 1)'$ and $(\bs{0}', 1)'$ of hyperplanes $H_{\bs{\beta}}$ and $H_h$, respectively. Then  $|\tan{\theta_{\bs{\beta}}}|=\|\bs{\beta}\|$ and (see Figure \ref{fig-4-proof}) $$|\bs{w}'_i\bs{\beta}|>\delta|\tan{\theta_{\bs{\beta}}}|=\delta\|\bs{\beta}\|>(1+\eta)\sqrt{M}.$$
 Now we have
\begin{align}
|r_i(\bs{\beta})|=|\bs{w}'_i\bs{\beta}-y_i| 
&\geq\big||\bs{w}'_i\bs{\beta}|-|y_i|\big| 
> (1+\eta)\sqrt{M}-|y_i|.
\end{align}
\vspace*{-5mm}
Therefore, \vs
\begin{align}
\sum_{j\in I(\bs{\beta})}r^2_j(\bs{\beta})\geq r^2_i(\bs{\beta}) 
&>\Big((1+\eta)\sqrt{M}-|y_i|\Big)^2  
 \geq \Big((1+\eta)\sqrt{M}-\sqrt{M}\Big)^2 \nonumber\\[1ex]
&=\eta^2M 
>M=\sum_{j\in I(\mb{0})}r^2_j(\bs{0}). \nonumber
\end{align}

\noin
That is, we have certified (\ref{proof.inequality}). \hfill \pend

\vs
\noindent
\tb{Proof of theorem 3.1}
\vs
\tb{Case A: $p=1$}. The problem becomes an estimation of a location parameter ${\beta_1}$ (the intercept term in the model $y_i=\beta_1+e_i$).  The solution is the depth trimmed mean based on ${y_i, i\in N}$, which has the RBP as claimed (see Wu and Zuo (2009)).
\vskip 3mm
\tb{Case B: $p>1$}.\vs
(i) \tb{First}, \emph{{we show that}} \emph{$m=\lfloor{n}/{2}\rfloor-p+2$ points are enough to breakdown $\widehat{\bs{\beta}}^n_{lst}$}.
 Recall the definition of $\widehat{\bs{\beta}}^n_{lst}$. One has
\begin{eqnarray}
\widehat{\bs{\beta}}_{lst}(\mb{Z}^{(n)}, \alpha)&=&\arg\min_{\bs{\beta}\in \R^p}Q(\bs{Z}^{(n)}, \bs{\beta}, \alpha)\nonumber  \\[2ex]
&=&\arg\min_{\bs{\beta}\in \R^p}\sum_{i=1}^{n}r_i^2\mathds{1}\left( \frac{|r_i-m(\bs{Z}^{(n)},\bs{\beta})|}{\sigma(\bs{Z}^{(n)},\bs{\beta})}\leq \alpha\right).
\label{T*-bp-proof.eqn}
\end{eqnarray}

\vskip 3mm
Select $p-1$ points from $\bs{Z}^{(n)}=\{ (\bs{x_i'}, y_i)'\}$. $(\bs{w}'_i, y_i)$, together with the origin, form a $(p-1)$-dimensional
subspace (hyperline) $L_h$ in the $(p+1)$-dimensional space of $(\bs{w}', y)'$.\vskip 3mm
 Construct a non-vertical hyperplane $H$ through $L_h$ (that is, it is not perpendicular to the horizontal hyperplane $y=0$). Let $\bs{\beta}$ be determined by the hyperplane $H$ through $y=\bs{w'}\bs{\beta}$.\vskip 3mm

 We can tilt the hyperplane $H$ so that it approaches its ultimate vertical position. Meanwhile, we put all the
 $m$ contaminating points onto this hyperplane $H$ so that it contains no less than $m+(p-1)=\lfloor {n/2}\rfloor+1$
 observations. Call the resulting contaminated sample by $\bs{Z}^{(n)}_m$.
 Therefore the majority of $r_i=y_i- \bs{w}'_i\bs{\beta}$ will now be zero. Therefore, $\sigma(\mb{Z}^{(n)}, \bs{\beta})$, in this case, is defined to be one.
 \vskip 3mm
When $H$ approaches  its ultimate vertical position,  $\|\bs{\beta}\|\to \infty$ (for the reasoning, see the \tb{case (II)} of the proof of Theorem 2.2) and $r_i$ for points $(\bs{w}'_i, y_i))'$ not on the $H$ will also approach $\infty$.
This implies that this $\bs{\beta}$ is the solution for $\widehat{\bs{\beta}}^n_{lst}$ at this contaminated data $\bs{Z}^{(n)}_m$ since it attains the minimum possible value (zero) on the RHS of (\ref{lst.eqn}).
That is, $m=\lfloor{n}/{2}\rfloor-p+2$ contaminating points are enough to break down $\widehat{\bs{\beta}}^n_{lst}$.  \vskip 3mm

(ii) \tb{Second},\emph{ we now show that}\emph{ $m=\lfloor{n}/{2}\rfloor-p+1$ points are not enough to break down $\widehat{\bs{\beta}}^n_{lst}$.}
Let $\bs{Z}^{(n)}_m$ be an arbitrary contaminated sample and $\bs{\beta_c}:=\widehat{\bs{\beta}}_{lst}(\bs{Z}^{(n)}_m, \alpha)$ and $\bs{\beta_o}=\widehat{\bs{\beta}}_{lst}(\bs{Z}^{(n)},\alpha)$,  where $\bs{Z}^{(n)}=\{\bs{Z}_i\}=\{ (\bs{x_i'}, y_i)'\}$ are uncontaminated original points. 
 Assume that $\bs{\beta_c}\neq \bs{\beta_o}$ (Otherwise, we are done). It suffices to show that $\|\bs{\beta_c}-\bs{\beta_o}\|$ is bounded.\vskip 3mm
 Note that since $n-m= \lfloor(n+1)/2\rfloor+p-1$, then both $m$ and $\sigma$ in respective (\ref{med.eqn}) and (\ref{mad.eqn})  are bounded for both contaminated $\bs{Z}^{(n)}_m$ and $\bs{\beta}_c$ and original $\bs{Z}^{(n)}$ and $\bs{\beta}_o$.
~Define
\bee
\delta &=&\frac{1}{2}\inf~\big\{\tau>0; ~\mbox{$\exists$ a $(p-1)$-dimensional subspace  $L$ of ($y=0$) such} \\
& & \mbox{that ${L}^{\tau}$ contains at least $p$ of uncontaminated  $(1,\bs{x'_i})$ from $\bs{Z}^{(n)}$}\big\},
\ene
where $L^{\tau}$ is the set of all points $\bs{w'}$ that have the distance to $L$ no greater than $\tau$.
Since $\bs{Z}^{(n)}$ is in general position,  $\delta>0$.\vskip 3mm
Let $H_o$ and $H_c$ be the hyperplanes determined by $y=\bs{w}'\boldsymbol{ \beta_o}$ and $y=\bs{w}'\boldsymbol{\beta_c}$, respectively, and $M=\max_{i}\{|y_i-\bs{w_i'}{\bs{\beta}}_o|\}$ for all original $y_i$ and $\bs{x_i}$ in $Z^{(n)}$. 
Since $\bs{\beta_o}\neq \bs{\beta_c}$, then $H_o\neq H_c$.\vskip 3mm

\textbf{{(I)}} \textbf{Assume that $H_o$ and $H_c$ are not parallel}.
Denote the vertical projection of the intersection
$H_o\cap H_c$ to the horizontal hyperplane $y=0$ by $L_{vp}(H_o\cap H_c)$, then it is $(p-1)$-dimensional.
By the definition of $\delta$, there are at most $p-1$ of  uncontaminated points of ${\bs{w}_i}=(1, \bs{x}'_i)'$ from the original $\{{\bs{Z}_i}, i=1,\cdots, n\}$ within $L_{vp}^{\delta}(H_o\cap H_c)$. Denote the set of all these  possible  $\bs{w}_i$ (at most $p-1$) by $S_{cap}$ and $|S_{cap}|=n_{cap}\leq (p-1)$.
 Denote the set of all remaining uncontaminated ${\bs{Z}_i}$  from the original $\{{\bs{Z}_i}, i=1,\cdots, n\}$ by $S_r$ and the set of all such $i$ as $J$, then there are at least $n- m-n_{cap}\geq n-\lfloor{n/2}\rfloor=\lfloor{(n+1)/2}\rfloor$ such $\bs{Z}_i$ in $S_r$.
\vskip 3mm

For each $(\bs{w'_i}, y_i)'$ with $i\in J$, construct a two-dimensional vertical plane $P_i$ that goes through $(\bs{w'_i}, y_i)'$ and $(\bs{w'_i}, y_i+1)'$ and is perpendicular to  $L_{vp}(H_o\cap H_c)$ (see Figure \ref{fig-4-proof} and/or Figure 16 of RL87). Denote the angle formed by $H_o$ and the horizontal line in
$P_i$  by $\alpha_o\in(-\pi/2, \pi/2)$, similarly by $\alpha_c$ for $H_c$ and $P_i$. They are essentially the  angles formed between $H_o$ and $H_c$ with the horizontal hyperplane $y=0$, respectively.
 \vskip 3mm
 We see that for $i\in J$ and each  $(\bs{w'_i}, y_i)'$,
$|\bs{w'_i}\bs{\beta_o}|>\delta|\tan(\alpha_o)|$ and $|\bs{w'_i}\bs{\beta_c}|>\delta|\tan(\alpha_c)|$ (see Figure \ref{fig-4-proof} or Figure 16 of RL87 of a geographical illustration for better understanding) and $\|\bs{\beta_o}\|=|\tan(\alpha_o)|$ and $\|\bs{\beta_c}\|=|\tan(\alpha_c)|$.
\vskip 3mm

 Now for each $i\in J$, 
 denote $r^o_i: =(y_i-\bs{w'_i}\bs{\beta_o})$ 
  and $r^c_i: =(y_i-\bs{w'_i}\bs{\beta_c})$. 
 For 
 any $i\in J$,
 it follows that (see Figure \ref{fig-4-proof} or Figure 16 of RL87)
 \bee
 |r^o_i-r^c_i|&=&\big|{\bs{w'_i}\bs{\beta_o}-\bs{w'_i}\bs{\beta_c}}\big|
 ~~>~~ {\delta |\tan(\alpha_o)-\tan(\alpha_c)|}
 ~~\geq~~ {\delta \big| |\tan(\alpha_o)|-|\tan(\alpha_c)|  \big|}\\[1ex]
 &=&{\delta\big|\|\bs{\beta_o}\|-\|\bs{\beta_c}\|\big|}
~~\geq~~ {\delta\big|\|\bs{\beta_o}-\bs{\beta_c}\|-2\|\bs{\beta_o}\|\big|}
 \ene
 \vskip 3mm

Let $M_1:=|m(\mb{Z}^{(n)}_m,\bs{\beta}_c)|+\alpha\sigma(\mb{Z}^{(n)}_m,\bs{\beta}_c)$, which is obviously bounded. Then it is obvious that
\begin{align}
Q(\mb{Z}^{(n)}_m, \bs{\beta}_c, \alpha)= \sum_{i\in I(\bs{\beta}_c)}(r^c_i)^2 \mathds{1}\left( \frac{|r^c_i-m(\bs{Z}^{(n)}_m,\bs{\beta})|}{\sigma(\bs{Z}^{(n)}_m,\bs{\beta})}\leq \alpha\right)\leq I(\bs{\beta}_c)M_1^2, \label{inequality.eqn}
\end{align}
\vs

If we assume that $\|\bs{\beta_o}-\bs{\beta_c}\|\geq 2\|\bs{\beta_o}\|+(M_1\sqrt{I(\bs{\beta}_c)}+M)\big/\delta$, 
then by the inequality above we have for $i\in J$ 
$$
|r^o_i-r^c_i|> {\delta\big|\|\bs{\beta_o}-\bs{\beta_c}\|-2\|\bs{\beta_o}\|\big|}
\geq M_1\sqrt{I(\bs{\beta}_c)}+M,
$$
which implies that for any $i\in J$, 
\[ |r^c_i|\geq |r^o_i-r^c_i|-|r^o_i| >  M_1\sqrt{I(\bs{\beta}_c)}+M-M = M_1\sqrt{I(\bs{\beta}_c)}.  
\]
\vs
Notice that $|J|\geq \lfloor(n+1)/2\rfloor$ which implies that there is at least one $i_0\in J$ that belongs to $I(\bs{\beta}_c)$ in light of Lemma 2.1.
Therefore
\bee
Q(\mb{Z}^{(n)}_m, \bs{\beta}_c, \alpha)&=&\sum_{i\in I(\bs{\beta}_c)}(r^c_i)^2 \mathds{1}\left( \frac{|r^c_i-m(\bs{Z}^{(n)}_m,\bs{\beta})|}{\sigma(\bs{Z}^{(n)}_m,\bs{\beta})}\leq \alpha\right)\\[1ex]
&\geq& (r^c_{i_0})^2> I(\bs{\beta}_c)M_1^2,
\ene
which contradicts (\ref{inequality.eqn}). That is, $\|\bs{\beta_o}-\bs{\beta_c}\| \left(< 2\|\bs{\beta_o}\|+(M_1\sqrt{I(\bs{\beta}_c)}+M)\big/\delta\right)$ is bounded.\vs

\textbf{{(II)}} \textbf{Assume that $H_o$ and $H_c$ are parallel}. That is,
$\bs{\beta_c}=\rho\bs{\beta_o}$. We claim that $\|\bs{\beta_c}-\bs{\beta_o}\|$ is bounded. If $\rho$ is finite or $\|\bs{\beta}_o\|=0$, then $\|\bs{\beta_c}-\bs{\beta_o}\|$ is automatically bounded. We are done.  Otherwise, consider the case that $\bs{\beta}_o\not =0$ and $|\rho| \to\infty$. 
\vs
\tb{(A)} \textbf{Assume that $H_o$ is not parallel to $y=0$}.\vskip 3mm
The proof is very similar to part \tb{(I)}.
Denote the intersection of $H_c$ and the horizontal hyperplane $y=0$: $H_c\cap\{y=0\}$ by $L_c$. Then
$L^\delta_c$ contains at most $p-1$ uncontaminated points from $\{\bs{Z}^{(n)}\}$. Denote the set of all the remaining uncontaminated points in $\{\bs{Z}^{(n)}\}$ as $S_r$. Hence $|S_r|\geq n-m-(p-1)\geq \lfloor(n+1/2\rfloor$.
Denote again by $J$ the set of all $i$ such that $\bs{Z}_i\in S_r$. Again let the angle between $H_c$ and $y=0$ be $\alpha_c$, then it is seen that $\|\bs{\beta_c}\|=|\tan(\alpha_c)|$ and $|\bs{w'_i}\mb{\bs{\beta_c}}|> \delta
|\tan(\alpha_c)|$ for any $i\in J$.
\vskip 3mm

Note that for $i\in J$, ~
$r^c_i=(y_i-\bs{w_i'}\mb{\bs{\beta_c}}).$ 
 Write $M_y=\max_{i\in J}|y_i|$. It follows that for $i\in J$
\[
\big|r^c_i\big|\geq \big||\mb{w'_i}\bs{\beta_c}|-|y_i|\big|\geq |~\delta|\tan(\alpha_c)|-M_y|.
\]
Since $|S_r|\geq \lfloor(n+1/2\rfloor$, then $M_1:=|m(\mb{Z}^{(n)}_m,\bs{\beta}_c)|+\alpha\sigma(\mb{Z}^{(n)}_m,\bs{\beta}_c)$ is obviously bounded (see reasing in (I) above) and
\begin{align}
Q(\mb{Z}^{(n)}_m, \bs{\beta}_c, \alpha)= \sum_{i\in I(\bs{\beta}_c)}(r^c_i)^2 \mathds{1}\left( \frac{|r^c_i-m(\bs{Z}^{(n)}_m,\bs{\beta})|}{\sigma(\bs{Z}^{(n)}_m,\bs{\beta})}\leq \alpha\right)\leq I(\bs{\beta}_c)M_1^2, \label{inequality-1.eqn}
\end{align}
Notice that $|J|\geq \lfloor(n+1)/2\rfloor$ which implies that there is at least one $i_0\in J$ that belongs to $I(\bs{\beta}_c)$ in light of Lemma 2.1.
Therefore
\bee
Q(\mb{Z}^{(n)}_m, \bs{\beta}_c, \alpha)&=&\sum_{i\in I(\bs{\beta}_c)}(r^c_i)^2 \mathds{1}\left( \frac{|r^c_i-m(\bs{Z}^{(n)}_m,\bs{\beta})|}{\sigma(\bs{Z}^{(n)}_m,\bs{\beta})}\leq \alpha\right)\\[1ex]
&\geq& (r^c_{i_0})^2> ~(\delta|\tan(\alpha_c)|-M_y)^2= (\delta|\rho|\|\bs{\beta_o}\|-M_y)^2
\ene
Since 
 $|\rho|$ could be arbitrarily large, then the above inequality contradicts  (\ref{inequality-1.eqn}).

 \vskip 3mm

\tb{(B)} \textbf{Assume that $H_o$ is parallel to $y=0$}.
Then, it means that $\bs{\beta_c}=\rho\bs{\beta_o}=(\rho\beta_{o1},0,\cdots, 0)$. Assume that $\beta_{o1}\neq 0$. Otherwise, we are done.  Now we can repeat the argument above since $n-m= (p-1)+\lfloor(n+1)/2\rfloor$.
Let $A$ be the set of all uncontaminated points from $\mb{Z}^{(n)}$, then $|A|=n-m=(p-1)+\lfloor(n+1)/2\rfloor$. Let $J$ be the set of all $i$ such that $\bs{Z}_i\in A$ and $M_y=\max_{i\in J}|y_i|$,
then $M_1:=|m(\mb{Z}^{(n)}_m,\bs{\beta}_c)|+\alpha\sigma(\mb{Z}^{(n)}_m,\bs{\beta}_c)$ is obvious bounded.
We still have
\begin{align}
Q(\mb{Z}^{(n)}_m, \bs{\beta}_c, \alpha)= \sum_{i\in I(\bs{\beta}_c)}(r^c_i)^2 \mathds{1}\left( \frac{|r^c_i-m(\bs{Z}^{(n)_m},\bs{\beta})|}{\sigma(\bs{Z}^{(n)_m},\bs{\beta})}\leq \alpha\right)\leq I(\bs{\beta}_c)M_1^2, \label{inequality-2.eqn}
\end{align}
 On the one hand  we have that for $i\in J$
 $$
 |r_i^c|=|\bs{w}'_i\bs{\beta}_c-y_i|\geq \big||\bs{w}'_i\bs{\beta}_c|-|y_i|\big|\geq\big||\rho||\beta_{o1}|-M_y\big|,
 $$
 which implies that $(r^c_i)^2$ becomes unbounded when $\rho \to \infty$. Since there is at least one $i_0\in J$ that belongs to $I(\bs{\beta}_c)$ in light of Lemma 2.1, now we have
 \bee
Q(\mb{Z}^{(n)}_m, \bs{\beta}_c, \alpha)&=&\sum_{i\in I(\bs{\beta}_c)}(r^c_i)^2 \mathds{1}\left( \frac{|r^c_i-m(\bs{Z}^{(n)}_m,\bs{\beta})|}{\sigma(\bs{Z}^{(n)}_m,\bs{\beta})}\leq \alpha\right)\\[1ex]
&\geq& (r^c_{i_0})^2\geq ~(|\rho||\beta_{o1}|-M_y)^2\to \infty ~~ (\mbox{as $\rho\to \infty$}),
\ene
 which contradicts to (\ref{inequality-2.eqn}).

That is, $m$ contaminating points are not enough to breakdown $\widehat{\bs{\beta}}^n_{lst}$ since $\|\bs{\beta}_o-\bs{\beta}_c\|$ remains bounded.
 \hfill \pend
\vs
\noin
\tb{Remark A.1}
\vs
 Parallel cases considered in the proofs of Theorems 2.2 and 3.1 (often missed the related discussions in the literature) are important. This is especially true in the latter case since one can not afford to miss the parallel cases when considering the all possibilities of contamination. \hfill \pend
\vs
\noindent
\tb{Proof of Lemma 3.2}
\vs
\noin
Denote the integrand in (\ref{Q.eqn}) as $G(\beta):=(y-\bs{w}'\bs{\beta})^2\mathds{1}\left( \frac{|y-\bs{w}'\bs{\beta}-m|}{\sigma}\leq \alpha\right)$ for a given point $(\bs{x}', y)\in \R^p$.
Write $G(\beta):=(y-\bs{w}'\bs{\beta})^2\big(1-\mathds{1}\left({|y-\bs{w}'\bs{\beta}-m|}\big/{\sigma}\leq \alpha\right)\big)$.
\vs
\tb{(i)} By the strictly non-flatness of $F_r$ around $m$ and $\sigma$, we have the continuity of the $m(\bs{\beta}$ and $\sigma(\bs{\beta})$. Consequently, $G(\bs{\beta})$ is obvious continuous in $\bs{\beta}\in \R^p$. Hence, $Q(\bs{\beta})$ is continuous in
$\bs{\beta}\in \R^p$.
\vs
\tb{(ii)} For arbitrary points $(\bs{x}', y)$ and $\bs{\beta}$ in $\R^p$ and fixed distribution $F_r$, there are three cases for consideration: (a) $|y-\bs{w}'\bs{\beta}-m|\big/{\sigma}< \alpha$
(b) $|y-\bs{w}'\bs{\beta}-m|\big/{\sigma}> \alpha$ and (c) $|y-\bs{w}'\bs{\beta}-m|\big/{\sigma}=\alpha$. Case (c) happens with probability zero, we thus skip this case and treat (a) and (b) only.
By the continuity in $\bs{\beta}$, there is a small neighborhood of $\bs{\beta}$: $B(\bs{\beta}, \delta)$, centered at $\bs{\beta}$ with radius $\delta>0$ such that (a) (or (b)) holds for all $\bs{\beta} \in B(\bs{\beta}, \delta)$. This implies that 
$$
\frac{\partial}{\partial \bs{\beta}}\mathds{1}\left( \frac{|y-\bs{w}'\bs{\beta}-m|}{\sigma}\leq \alpha\right)=\bs{0},
$$
and
\[
\frac{\partial}{\partial \bs{\beta}}G(\bs{\beta})=-2(y-\bs{w}'\bs{\beta})\bs{w}\mathds{1}\left(\frac{|y-\bs{w}'\bs{\beta}-m|}{\sigma}\leq \alpha)\right),
\]
Hence, we have that
\[
\frac{\partial^2}{\partial \bs{\beta}^2}G(\bs{\beta})=2\bs{w}\bs{w}'\mathds{1}\left(\frac{|y-\bs{w}'\bs{\beta}-m|}{\sigma}\leq \alpha)\right),
\]
Note that $G(\bs{\beta})$ is uniformly bounded over $\bs{\beta} \in \R^p$, then by the Lebesgue dominated convergence theorem, the desired result follows.
\vs
\tb{(iii)} The convexity follows from the twice differentiability and the positive semidefinte of the second order derivative of $Q(\bs{\beta})$ and the strict convexity follows from the invertibility of Hessian matrix.
\hfill \pend
\vs
\noindent
\tb{Proof of Theorem 3.2}
\vs
\noin
 We will treat $\bs{\beta}_{lts}(F_{(\bs{x};, y)},\alpha)$ , the counterpart for $\bs{\beta}_{lts}(F_{\varepsilon}(\bs{z}),\alpha)$  can be treated analogously.
\vs
\tb{(i)} Existence follows from the positive smidefinite of the Hessian matrix (see proof of (ii) of Lemma 3.2) and the convexity of $Q(\bs{\beta})$.\vs
\tb{(ii)} The equation follows from  the Lebesgue dominated convergence theorem, the differentiability and the first order derivative of $Q(\bs{\beta})$ given in the proof (ii) of Lemma 3.2. \vs
\tb{(iii)} The uniqueness follows from  the Lebesgue dominated convergence theorem, the positive definite of the Hessian matrix based on the given condition (invertibility).
 \hfill \pend
\vs
\noin
\vs
\noindent
\tb{Remarks 3.2}
\vs
\tb{(I)} Generally, the influence function  for a regression estimator when $p>1$  is not often provided in the literature (exceptions including Zuo (2021b) for the projection regression median, and \"{O}llerer, et al (2015) for the penalized regression estimators. 
In the latter case for the spare LTS, it is still restricted to $p=1$ and $x$ and $e$ are independent and normally distributed, though). In the location setting ($p=1$) the IF of the LTS estimator has been given in Tableman (1994). 
In this special case ($p=1$) in our model (\ref{model.eqn}), we have a location problem for the $\beta_{01}$ and the IF was given in Wu and Zuo (2019) and is bounded.
\vs
\tb{(II)} If setting $\alpha \to \infty$, then one immediately obtains the influence function for  LS estimating functional, $\bs{\beta}_{ls}$, which is with $\mb{z}_0=(\mb{s'}_0, t_0)'\in \R^{p}$
$$
\mbox{IF}(\mb{z}_0; \bs{\beta}_{ls}, F_{(\bs{x'}, y)})=(E(\bs{w}\bs{w}'))^{-1}(1,\mb{s}'_0)'(t_0-(1,\mb{s}_0')\bs{\beta}_{ls}).
$$
Of course, assuming that the inverse exists. 
Obviously, one can follow the approach in the theorem to obtain the IF for LTS in the case $p>1$.
\vs
\tb{(III)} When the depth of the residual of the contaminating point $\mb{z}'_0=(\mb{s}'_0, t_0)$ with respect to the $\bs{\beta}_{lst}$ is larger than $\alpha$, then the point mass contamination does not affect at all the functional $\bs{\beta}_{lst}$ with its influence function remaining bounded.
 It, 
  unfortunately, might be unbounded (in $p>1$ case), sharing the same drawback of that of LTS (in the $p=1$ case). The latter was shown in  \"{O}llerer, et al (2015)
  even in the simple regression case with  $x$ and $e$ are independent and normally distributed.
\hfill \pend
\vs
\noindent
\tb{Proof of theorem 3.3}

\noin
Insert $\bs{\beta}^{\varepsilon}_{lst} (\mb{z}_0):=\bs{\beta}_{lst}(F_{\varepsilon}(\mb{z}_0), \alpha)$ for $\bs{\beta}$ in (\ref{lts-contamination-estimation.eqn}) and take derivative with respect to $\varepsilon$ and let $\varepsilon \to 0$, we obtain (in light of dominated convergence theorem)
\begin{align}
&\left(\int \frac{\partial}{\partial{\bs{\beta}^{\varepsilon}_{lst}(\mb{z}_0)}}\left(r(\bs{\beta}^{\varepsilon}_{lst} (\mb{z}_0))\bs{v}\mathds{1}(\bs{\beta}^{\varepsilon}_{lst} (\mb{z}_0), F_{\varepsilon}(\mb{z}_0)
 \right)\Big|_{\varepsilon = 0}\!\!\!
dF_{(\bs{x'}, y)}\!\right)\!\dot{\bs{\beta}}_{lst}(\mb{z}_0, F_{(\bs{x'}, y)})\nonumber\\[3ex]
+& I_2 -I_3= \mb{0}, \label{if-1.eqn}
\end{align}
where $r(\bs{\beta})=y-\bs{w}'\bs{\beta}$,  $\mathds{1}(\bs{\beta}, G)=\mathds{1}\left({|(y-\bs{w}'\bs{\beta})-m(G)|}\big/{\sigma(G)}\leq \alpha\right)$, and
\begin{align}
I_2=& \int (r(\bs{\beta}_{lst}(F_{(\bs{x'}, y)},\alpha))\bs{v}
\mathds{1}(\bs{\beta}_{lst}(F_{(\bs{x'}, y)},\alpha), F_{(\bs{x'}, y)}),\nonumber\\[1ex]
I_3=&\int (r(\bs{\beta}_{lst}(F_{(\bs{x'}, y)},\alpha))\bs{w}
\mathds{1}(\bs{\beta}_{lst}(F_{(\bs{x'}, y)},\alpha), F_{(\bs{x'}, y)})dF_{(\bs{x'}, y)}.\nonumber
\end{align}
Denote by $I_1$ for the first term on the LHS of the above first equation. We have $I_1+I_2-I_3=0$, and 
\bee
I_2-I_3&
=&(t_0-(1, \mb{s}'_0)\bs{\beta}_{lst}(F_{(\bs{x'}, y)},\alpha))(1, \mb{s}'_0)'
\mathds{1}\left( \frac{|(t_0-(1, \mb{s}'_0)\bs{\beta}_{lst}(F_{(\bs{x'}, y)}, \alpha)-m|}{\sigma}\leq \alpha\right),\\[0ex]
\ene
where the equality follows from (\ref{lst-estimation.eqn}) (i.e. $I_3=\bs{0}$). The RHS of the last display is:
\bee
&=&\left\{
\begin{array}{ll}
\mb{0}, &\hspace*{-10mm}\mbox{if $t_0-(1, \mb{s}'_0)\bs{\beta}_{lst} \not \in [m(\bs{\beta}_{lst})-\alpha\sigma(\bs{\beta}_{lst}),~~ m(\bs{\beta}_{lst})+\alpha\sigma(\bs{\beta}_{lst})]$},\\[2ex]
(t_0-(1, \mb{s}'_0)\bs{\beta}_{lst})(1, \mb{s}'_0)',& \mbox{~~otherwise},
\end{array}
\right.
\ene

Now we focus on the $I_1$ and especially its integrand. Denote the latter by $I_4$. We have
\begin{align}
I_4&=\frac{\partial}{\partial{\bs{\beta}^{\varepsilon}_{lst} (\mb{z}_0)}}\left((y-\bs{w}'\bs{\beta}^{\varepsilon}_{lst} (\mb{z}_0))\bs{w}\mathds{1}\left( \frac{|(y-\bs{w}'\bs{\beta}^{\varepsilon}_{lst} (\mb{z}_0))-m_{\varepsilon}(\mb{z}_0)|}{\sigma_{\varepsilon}(\mb{z}_0)}\leq \alpha\right)\right)\Bigg|_{\varepsilon = 0}\nonumber\\[2ex]
&=\left(-\bs{w}\bs{w'}\mathds{1}\left( \frac{|(y-\bs{w}'\bs{\beta}^{\varepsilon}_{lst} (\mb{z}_0))-m_{\varepsilon}(\mb{z}_0)|}{\sigma_{\varepsilon}(\mb{z}_0)}\leq \alpha\right) \right)\Bigg|_{\varepsilon = 0}
\nonumber\\[2ex]
&+\left(
(y-\bs{w}'\bs{\beta}^{\varepsilon}_{lst} (\mb{z}_0))\bs{w}\frac{\partial}{\partial{\bs{\beta}^{\varepsilon}_{lst} (\mb{z}_0)}}\mathds{1}\left( \frac{|(y-\bs{w}'\bs{\beta}^{\varepsilon}_{lst} (\mb{z}_0))-m_{\varepsilon}(\mb{z}_0)|}{\sigma_{\varepsilon}(\mb{z}_0)}\leq \alpha\right)\right)\Bigg|_{\varepsilon = 0}. \nonumber
\end{align}
Hence
\begin{align}
I_4
&=-\bs{w}\bs{w'}\mathds{1}\left( \frac{|(y-\bs{w}'\bs{\beta}_{lst})-m(\bs{\beta}_{lst})|}{\sigma(\bs{\beta}_{lst})}\leq \alpha\right)\nonumber\\[2ex]
&+(y-\bs{w}'\bs{\beta}_{lst})\bs{w}\frac{\partial}{\partial{\bs{\beta}}}\mathds{1}\left( \frac{|(y-\bs{w}'\bs{\beta})-m(\bs{\beta})|}{\sigma(\bs{\beta})}\leq \alpha\right)\Bigg|_{\bs{\beta}=\bs{\beta}_{lst}}\nonumber\\[2ex]
&=-\bs{w}\bs{w'}\mathds{1}\left( \frac{|(y-\bs{w}'\bs{\beta}_{lst})-m(\bs{\beta}_{lst})|}{\sigma(\bs{\beta}_{lst})}\leq \alpha\right),\nonumber
\end{align}
where the last step follows from the proof of  Lemma 3.2.
\vs
Now we have in light of (\ref{if-1.eqn})
\bee
\left(\int(-I_4)dF_{(\bs{x'}, y)}\right)\dot{\bs{\beta}}_{lst}(\mb{z}_0, F_{(\bs{x'}, y)})=I_2.
\ene
The desired result follows. \hfill \pend

\vs
\noindent
\tb{Proof of lemma 4.2} \vs
It suffices to establish (a), (b) follows straightforwardly. Put $m_{\sup}=\sup_{\bs{\beta} \in\Theta }m(F_{y-\bs{w'}\bs{\beta}})$, $m_{\inf}=\inf_{\bs{\beta}\in\Theta }m(F_{y-\bs{w'}\bs{\beta}})$, and $\sigma_{\sup}=\sup_{\bs{\beta}\in\Theta }\sigma(F_{y-\bs{w'}\bs{\beta}})$,
by continuity in $\bs{\beta}$ and boundedness of $\Theta$, all are finite numbers. Define two classes of functions for a fixed $\alpha$, $m_{\sup}$, $m_{\inf}$, and $\sigma_{\sup}$ with $r(\bs{\beta})=y-\bs{w}'\bs{\beta}$
\bee
\mathscr{F}_1(\bs{\beta}):&=&\left\{f(\bs{x}, y, \bs{\beta})=(r(\bs{\beta}))^2\mathds{1}\left( \frac{|r(\bs{\beta})-m({F_R})|}{\sigma(F_R)}\leq \alpha\right), \bs{\beta}\in\Theta \right\}, \\[2ex]
\mathscr{F}_2(\bs{\beta}):&=&\left\{f(\bs{x}, y, \bs{\beta})=(r(\bs{\beta}))^2\mathds{1}\left(m_{\inf}-\alpha\sigma_{\sup} \leq r(\bs{\beta})\leq m_{\sup}+\alpha\sigma_{\sup} \right), \bs{\beta}\in\Theta \right\}. 
\ene
Obviously, $\mathscr{F}_1(\bs{\beta})\subset \mathscr{F}_2(\bs{\beta})$. Following the notation of Pollard (1984)(P84), we have for any $\bs{\beta} \in \Theta$, $$Q(F^n_{\mb{Z}},\bs{\beta})-Q(F_{\mb{Z}},\bs{\beta})= P_nf(\bs{x}, y, \bs{\beta})-Pf(\bs{x}, y, \bs{\beta}):=P_n f-Pf,$$
where $f:=f(\bs{x}, y, \bs{\beta})\in \mathscr{F}_1(\bs{\beta})$ (hereafter for consistency we assume that there is a factor $\frac{1}{n}$ in the RHS of (\ref{objective.eqn}). This will not affect the minimization or all previous discussions). And
\be
\sup_{\bs{\beta}\in\Theta}|Q(F^n_{\mb{Z}},\bs{\beta})-Q(F_{\mb{Z}},\bs{\beta})|=\sup_{f\in \mathscr{F}_1(\bs{\beta})}|P_n f-Pf|\leq \sup_{f\in \mathscr{F}_2(\bs{\beta})}|P_n f-Pf|. \label{inequality-4.eqn}
\ee
It suffices to show the  most right hand side equals to $o(1)$ a.s. (cf, supplement of Zuo(2020) for this part of proof).\vs

To achieve that, we invoke Theorem 24 of II.5 of P84. First $\mathscr{F}_2(\bs{\beta})$ is a permissible class of functions with an envelop
$F=(m_{\sup}+\alpha\sigma_{\sup})^2$. Second, to verify the logarithm of the covering number is $o_p(n)$, by Theorem 25 of II.5 of P84, it suffices to show that
the graphs of functions in $\mathscr{F}_2(\bs{\beta})$  have only polynomial discrimination (for related concepts, cf P84), also see Example 26 of II.5 of P84 (page 29) and Example 18 of VII.4 of P84 (page 153).\vs

The graph of a real-valued function $f$ on a set $S$ is defined as the subset (see page 27 of P84)
$$G_f = \{(s, t): 0\leq t \leq f(s) ~\mbox{or}~ f(s)\leq t \leq 0, s \in S \}.$$
\vs
The graph of a function in $\mathscr{F}_2(\bs{\beta})$ contains a point $(\mb{x(\omega)}, y(\omega), t)$  if and only if
$0\leq t\leq f(\bs{x}, y, \bs{\beta})$ or $ f(\bs{x},y, \bs{\beta}) \leq t\leq 0$. The latter case could be excluded since the function is always nonnegative (and equals $0$ case covered by the former case). The former case happens if and only if
$0\leq\sqrt{t}\leq y-\bs{w'}\bs{\beta}$.\vs Given a collection of $n$ points,  the graph of a function in $\mathscr{F}_2(\bs{\beta})$ picks out only points that belong  
to $\{\sqrt{t}\geq 0\}\cap \{y-\bs{\beta}'\bs{w}-\sqrt{t}\geq 0\}$. Given $n$ points $(\bs{x}_i, y_i, t_i)$ ($t_i\geq 0$), introduce $n$ new points $(\bs{x}_i, y_i, z_i):=(\bs{x}_i, y_i, \sqrt{t_i})$ in $\R^{p+1}$. On $\R^{p+1}$ define a vector space $\mathscr{G}$ of functions
$$g_{a, b, c}(\bs{x}, y, z)=\mb{a}'\bs{x}+by+cz,$$
where $a\in \R^p$, $b\in \R^1$, and $c\in \R^1$ and $\mathscr{G}:=\{g_{a, b, c}(\bs{x}, y, z)=\mb{a}'\bs{x}+by+cz, a\in \R^p, b\in \R^1, ~\mbox{and}~ c\in \R^1\}$ which is $\R^{p+1}$-dimensional vector space.
\vs

It is clear now that the graph of a function in $\mathscr{F}_2(\bs{\beta})$ picks out only points that belong to
 the sets of $\{g\geq 0\}$ for $g\in \mathscr{G}$. By Lemma 18 of II.4 of P84 (page 20), the graphs of functions in $\mathscr{F}_2(\bs{\beta})$  pick only polynomial numbers of subsets of $\{w_i:=(\bs{x}_i, y_i, z_i), i=1,\cdots, n\}$; those sets corresponding to
$g\in  \mathscr{G}$ with $a \in \{\bs{0},-\bs{\beta}\}$, $b\in \{0, 1\}$, and $c \in \{1, -1\}$  pick up even few subsets from $\{w_i, i=1,\cdots, n\}$. This in conjunction with Lemma 15 of II.4 of P84 (page 18), yields that
the graphs of functions in $\mathscr{F}_2(\bs{\beta})$  have only polynomial discrimination. \vs
By Theorem 24 of II.5 of P84 we have completed the proof.
\hfill \pend
\vs
\noindent
\tb{Proof of lemma 4.3} \vs
 Assume conversely that $\sup_{\bs{\beta}\in N^c_{\ep}(\bs{\eta})}D(\bs{\beta}; F_{\mb{Z}})=D(\bs{\eta}; F_{\mb{Z}})$. Then by the given conditions, there is a sequence of \emph{bounded} $\bs{\beta_j}$ ($j=0,1,\cdots$) in $N^c_\ep(\bs{\eta})$ such that   $\bs{\beta_j}\to \bs{\beta_0}\in N^c_\ep(\bs{\eta})$ and $D(\bs{\beta_j}; F_{\mb{Z}})\to D(\bs{\eta}; F_{\mb{Z}})$ as $j\to \infty$.  Note that $D(\bs{\eta}; F_{\mb{Z}})>D(\bs{\beta_0}; F_{\mb{Z}})$. The continuity of $D(\cdot; F_{\mb{Z}})$ now leads to a contradiction: for sufficiently large $j$, $D(\bs{\beta_j}; F_{\mb{Z}})\leq (D(\bs{\eta}; F_{\mb{Z}}) +D(\bs{\beta_0};F_{\mb{Z}}))/2< D(\bs{\eta}; F_{\mb{Z}})$.  This completes the proof. \hfill \pend
\vs
\noindent
\tb{Proof of theorem 4.3} \vs
For convenience of description, we write
\be\mathds{1}(\bs{\beta}, F_{r(\bs{\beta})}):= \mathds{1}\left( \frac{|y-\bs{w'}\bs{\beta}-m({F_{r(\bs{\beta})}})|}{\sigma(F_{r(\bs{\beta})})}\leq \alpha\right),\label{indicator.eqn}
\ee
where $r(\bs{\beta})=y-\bs{w'}\bs{\beta}$ and $m(F_{r(\bs{\beta})})$ and $\sigma (F_{r(\bs{\beta})})$ are the median and MAD of the distribution of $r(\bs{\beta})$.\vs Adding the derivative of $Q(\mb{Z}^{(n)},\bs{\beta},\alpha)$ with respect to $\bs{\beta}$ evaluated at $\bs{\beta}=\bs{\beta}_0$ to the both sides of equation (\ref{estimation.eqn}) and multiplying $1/(2\sqrt{n})$ we obtain
\begin{align}
&\frac{1}{\sqrt{n}}\sum_i(y_i-\bs{w}'_i\bs{\beta}_0)\bs{w}_i\mathds{1}(\bs{\beta}_0, F^n_{r(\bs{\beta}_0)})=\nonumber\\[1ex]
&\frac{1}{\sqrt{n}}\sum_{i}\bs{w}_i\bs{w'}_i(\widehat{\bs{\beta}}^n_{lst}-\bs{\beta}_0)\mathds{1}(\bs{\beta}_0, F^n_{r(\bs{\beta}_0)})-\frac{1}{\sqrt{n}}\sum_{i}r_i(\widehat{\bs{\beta}}^n_{lst})\bs{w}_i\Big[\mathds{1}(\widehat{\bs{\beta}}^n_{lst}, F^n_{r(\widehat{\bs{\beta}}^n_{lst})})-\mathds{1}(\bs{\beta}_0, F^n_{r(\bs{\beta}_0)})\Big]\nonumber, 
\end{align}
where $\mathds{1}(\bs{\beta}, F^n_{r(\bs{\beta})})$ has the same meaning as in (\ref{indicator.eqn}) except that the median and MAD are the sample version, respectively based on $\{y_i-\bs{w'}_i\bs{\beta}\}$.
For further simplicity, we write $\mathds{1}(\bs{\beta}, n)$ for $\mathds{1}(\bs{\beta}, F^n_{r(\bs{\beta})})$, and $I_0$ for the LHS of the equation above. Rewrite the RHS of the equation above, we have
\begin{align}
&\frac{1}{\sqrt{n}}\sum_i(y_i-\bs{w}'_i\bs{\beta}_0)\bs{w}_i\mathds{1}(\bs{\beta}_0, F^n_{r(\bs{\beta}_0)})=
\frac{1}{n}\sum_i\bs{w}_i\bs{w}'_i\mathds{1}(\bs{\beta}_0, n)\sqrt{n}(\widehat{\bs{\beta}}^n_{lst}-\bs{\beta}_0)\nonumber\\[1ex]
&+\frac{1}{n}\sum_i\bs{w}_i\bs{w}'_i\Big[\mathds{1}(\widehat{\bs{\beta}}^n_{lst}, n)-\mathds{1}(\bs{\beta}_0, n)\Big]\sqrt{n}(\widehat{\bs{\beta}}^n_{lst}-\bs{\beta}_0) 
-\frac{1}{\sqrt{n}}\sum_{i}e_i\bs{w}_i\Big[\mathds{1}(\widehat{\bs{\beta}}^n_{lst}, n)
-\mathds{1}(\bs{\beta}_0, n)\Big]& \nonumber
\end{align}
Denote the three terms on the RHS above as $I_1$, $I_2$, and $I_3$, respectively.  Now we have, based on the short notations,
$$I_0=I_1+I_2+I_3.$$
If we can show that $I_0=O_p(1)$, $I_1=(O_p(1)+o_p(1))\sqrt{n}(\widehat{\bs{\beta}}^n_{lst}-\bs{\beta}_0)$, $I_2= o_p(1)\sqrt{n}(\widehat{\bs{\beta}}^n_{lst}-\bs{\beta}_0)$, and $I_3=o_p(1)$,
then the desired result follows immediately. On the other hand, these results are established in Lemmas 4.4 and 4.5. This completes the proof.
\hfill \pend

\vs
\noin
\tb{Lemma 4.4} With the assumptions \tb{(A3)}-\tb{(A4)}, we have $$\frac{1}{\sqrt{n}}\sum_i(y_i-\bs{w}'_i\bs{\beta}_0)\bs{w}_i\mathds{1}(\bs{\beta}_0, F^n_{R(\bs{\beta}_0)})=O_p(1).$$
\vs
\noindent
\tb{Proof}: Notice that $y_i-\bs{w}'_i\bs{\beta}_0=e_i$. It suffices to show that
$$\frac{1}{\sqrt{n}}\sum_ie_i\bs{w}_i=O_p(1).$$
This however follows straightforwardly from the CLT and E$(e_i\bs{w}_i)=0$.
\hfill \pend
\vs
\noin
\tb{Lemma 4.5}   With the assumptions  \tb{(A0)}-\tb{(A4)}, we have
\begin{align}
\frac{1}{n}\sum_i\bs{w}_i\bs{w}'_i\mathds{1}(\bs{\beta}_0, n)\sqrt{n}(\widehat{\bs{\beta}}^n_{lst}-\bs{\beta}_0)&=(O_p(1)+o_p(1))\sqrt{n}(\widehat{\bs{\beta}}^n_{lst}-\bs{\beta}_0),\\[1ex]
\frac{1}{n}\sum_i\bs{w}_i\bs{w}'_i\Big[\mathds{1}(\widehat{\bs{\beta}}^n_{lst}, n)-\mathds{1}(\bs{\beta}_0, n)\Big]\sqrt{n}(\widehat{\bs{\beta}}^n_{lst}-\bs{\beta}_0)&= o_p(1)\sqrt{n}(\widehat{\bs{\beta}}^n_{lst}-\bs{\beta}_0),\\[1ex]
\frac{1}{\sqrt{n}}\sum_{i}e_i\bs{w}_i\Big[\mathds{1}(\widehat{\bs{\beta}}^n_{lst}, n)
-\mathds{1}(\bs{\beta}_0, n)\Big]&=o_p(1).
\end{align}

\noindent
\tb{Proof:}  
By theorems 4.1 and 4.2, we have that $\widehat{\bs{\beta}}^n_{lst}-\bs{\beta}_0=o(1)$ a.s. Furthermore, sample median $m(F^n_{r(\bs{\beta}_0)})$ converges to its popular version $m(F_{r(\bs{\beta}_0)})$ a.s. by Glivenko-Cantelli theorem, the continuity of the median functional (see page 7 of Pollard (1984) (P84)), and Theorem 2.3.1 of Serfling (1980), hence we have $$\mathds{1}(\bs{\beta}_0, n)= \mathds{1}(\bs{\beta}_0, F_{r(\bs{\beta}_0)})+o(1), a.s.~~~\mbox{and}~~~ \mathds{1}(\widehat{\bs{\beta}}^n_{lst}, n)
-\mathds{1}(\bs{\beta}_0, n)=o(1), ~a.s.~$$
In light of the CLT and by \tb{(A3)} and \tb{(A4)}, we have that
$$
\frac{1}{\sqrt{n}}\sum_{i}e_i\bs{w}_i=\sqrt{n} E(e\bs{w})+O_p(1)=O_p(1).
$$
Now in virtue of the LLN, we have that
$$
\frac{1}{n}\sum_i\bs{w}_i\bs{w}'_i=E(\bs{w}\bs{w'})+o_p(1).
$$
The last three displays lead to the desired results.
\hfill \pend
\vs
\noindent
\tb{Proof of Theorem 5.1} \vs
In order to apply the Lemma 5.1, we first realize that in our case, $\widehat{\bs{\beta}}^n_{lst}$ and $\bs{\beta}_{lst}$ correspond to $\tau_n$ and $t_0$ (assume, w.l.o.g. that $\bs{\beta}_{lts}=\mb{0}$ in light of regression equivariance); $\bs{\beta}$ and $\Theta$ correspond to $t$ and $T$; $f(\cdot, t):= f(\cdot, \cdot, \bs{\beta}, \alpha)$ and $\alpha$ is a fixed constant, where $f(\bs{x}, y, \bs{\beta}, \alpha)=(y-\bs{w}'\bs{\beta})^2\mathds{1}(F_{(\bs{x}', y)}, \bs{\beta},\alpha)$ and $\mathds{1} (F_{(\bs{x'}, y)}, \bs{\beta}, \alpha):=\mathds{1}\left( \frac{|y-\bs{w}'\bs{\beta}-\mu(F_r)|}{\sigma(F_r)}\leq \alpha\right)$.
$r=y-\bs{w}'\bs{\beta}$.
In our case,
\[
\nabla(\bs{x}, y, \bs{\beta}, \alpha)=\frac{\partial}{\partial \bs{\beta}} f(\bs{x}, y, \bs{\beta}, \alpha)=2(y-\bs{w}'\bs{\beta})\bs{w}\mathds{1}(F_{(\bs{x}', y)}, \bs{\beta},\alpha).
\]
We will have to assume that $P(\nabla^2_i)=P(4(y-\bs{w}'\bs{\beta})^2{w}^2_i\mathds{1}(F_{(\bs{x}', y)}, \bs{\beta},\alpha)$ exists to meet (iv) of the lemma, where $i\in \{1,\cdots, p\}$ and $\bs{w}'=(w_1, \cdots, w_p)=(1, \bs{x}')$. It is readily seen that a sufficient condition for this assumption to hold is the existence of $P(x^2_i)$. In our case,
$V=2 P(\bs{w}\bs{w}'\mathds{1}(F_{(\bs{x}', y)}, \bs{\beta},\alpha)$, we will have to assume that it is invertible when $\bs{\beta}$ is replaced by $\bs{\beta}_{lst}$ (it is covered by the assumption in Theorem 3.2)  to meet (ii) of the lemma. In our case,
\[r(\cdot, t)=\left(\frac{\bs{\beta}'}{\|\bs{\beta}\|}V/2 \frac{\bs{\beta}}{\|\bs{\beta}\|} \right)\|\bs{\beta}\|.\]
We will assume that $\lambda_{min}$ and $\lambda_{max}$ are the minimum and maximum eigenvalues of positive semidefinite matrix $V$ overall $\bs{\beta}\in \Theta$ and a fixed $\alpha \geq 1$.%
\vs
Now to apply Lemma 5.1, we need to verify the five conditions, among them only (iii) and (v) need to be addressed, all others are satisfied trivially. For (iii), it holds automatically since our $\tau_n=\widehat{\bs{\beta}}^n_{lst}$ is defined to be the minimizer of  $F_n(t)$ over $t\in T (=\Theta)$.\vs
So the only condition that needs to be verified is the (v), the stochastic equicontinuity of $\{E_nr(\cdot, t)\}$ at $t_0$. For that, we will appeal to the Equicontinuity Lemma (VII.4 of P84, page 150).
To apply the Lemma, we will verify that the condition for the random covering numbers satisfy the uniformity condition. To that end, we look at the class of functions for a fixed $\alpha\geq 1$
\[
\mathscr{R}(\bs{\beta})=\left\{r(\cdot, \cdot, \alpha, \bs{\beta})=\left(\frac{\bs{\beta}'}{\|\bs{\beta}\|}V/2 \frac{\bs{\beta}}{\|\bs{\beta}\|} \right)\|\bs{\beta}\|:~ \bs{\beta}\in \Theta \right\}.
\]
Obviously, $\lambda_{max} r_0/2$ is an envelope for the class $\mathscr{R}$ in $\mathscr{L}^2(P)$, where $r_0$ is the radius of the ball $\Theta=B(\bs{\beta}_{lts}, r_0)$. We now show that the covering numbers of $\mathscr{R}$ are uniformly bounded, which amply suffices for the Equicontinuity Lemma.  For this, we will invoke Lemmas II.25 and II.36 of P84. 
 To apply Lemma II.25, we need to show that the graphs of functions in $\mathscr{R}$ have only polynomial discrimination.
 \vs
The graph of a real-valued function $f$ on a set $S$ is defined as the subset (see page 27 of P84 )
$$G_f = \{(s, t): 0\leq t \leq f(s) ~\mbox{or}~ f(s)\leq t \leq 0, s \in S \}.$$
\vs
  The graph of $r(\bs{x}, y, \alpha, \bs{\beta})$ contains a point $(\bs{x}, y, t)$, $t\geq 0$
  iff $\left(\frac{\bs{\beta}'}{\|\bs{\beta}\|}V/2 \frac{\bs{\beta}}{\|\bs{\beta}\|} \right)\|\bs{\beta}\| \geq t$ for all $\bs{\beta} \in \Theta$. 
 Equivalently, the graph of $r(\bs{x}, y, \alpha, \bs{\beta})$ contains a point $(\bs{x}, y, t)$, $t\geq 0$ if and only if $\lambda_{min}/2\|\bs{\beta}\|\geq t$.
For a collection of $n$ points $(\bs{x}'_i, y_i, t_i)$ with $t_i\geq 0$, the graph picks out those points satisfying $\lambda_{min}/2\|\bs{\beta}\|- t_i\geq 0$. Construct from
$(\bs{x}_i, y_i, t_i)$ a point $z_i=t_i$ in $\R$. On $\R$ define a vector space $\mathscr{G}$ of functions
\[
g_{a, b}(x)=ax+b,~~ a,~ b \in \R.
\]
By Lemma 18 of P84, the sets $\{g\geq 0\}$, for $g \in \mathscr{G}$, pick out only a polynomial number of subsets from $\{z_i\}$; those sets corresponding to functions in $\mathscr{G}$ with $a=-1$ and $b=\lambda_{min}/2\|\bs{\beta}\|$ pick out even fewer subsets from  $\{z_i\}$. Thus the graphs of functions in $\mathscr{R}$ have only    polynomial discrimination.
\hfill \pend
\vs
\vs
\noindent
\tb{Transformation in Section 5 before Corollary 5.1} ~
Assume the Cholesky decomposition of $\bs{\Sigma}$ in (\ref{ell.eqn}) yields a nonsingular lower triangular matrix $\bs{L}$ of the form
\[
\left(
\begin{array}{cc}
\bs{A} & \bs{0}\\
\bs{v}'& c
\end{array}
\right)
\]
with $\bs{\Sigma}=\bs{L}\bs{L}'$. Hence $\det(\bs{A})\neq 0\neq c$. Now transfer $(\bs{x}', y)$ to $(\bs{s}', t)$ with  $(\bs{s}', t)'=\bs{L}^{-1}((\bs{x}', y)'-\bs{\mu})$. It is readily seen that the distribution of $(\bs{s}', t)'$ follows
$E(g; \bs{0}, \bs{I_{p\times p}})$. \vs
Note that $(\bs{x}', y)'=\bs{L}(\bs{s}', t)'+(\bs{\mu}'_1, \mu_2)'$ with $\bs{\mu}=(\bs{\mu}'_1, \mu_2)'$. That is,
\begin{align}
\bs{x}&=\bs{A}\bs{s}+\bs{\mu}_1,\\[1ex]
y&=\bs{v}'\bs{s}+ct+\mu_2.
\end{align}
Equivalently,
\begin{align}
(1,\bs{s}')'&=\bs{B}^{-1}(1,\bs{x}')',   \label{x-transformation.eqn}\\[1ex]
t&=\frac{y-(1,\bs{s}')(\mu_2, \bs{v}')'}{c}, \label{y-transformation.eqn}
\end{align}
where
\[
\bs{B}=\begin{pmatrix}
1 &\bs{0}'\\
\bs{\mu}_1  &\bs{A}
\end{pmatrix}
,~~~~
\bs{B}^{-1}=
\begin{pmatrix}
1 &\bs{0}'\\
-\bs{A}^{-1}\bs{\mu}_1& \bs{A}^{-1}
\end{pmatrix},
\]
\vs
It is readily seen that (\ref{x-transformation.eqn}) is an affine transformation on $\bs{w}$ and (\ref{y-transformation.eqn}) is first an affine transformation on $\bs{w}$ then a regression transformation on $y$ followed by a scale transformation on $y$. In light of Theorem 2.4, we can assume hereafter, w.l.o.g. that $(\bs{x}', y)$ follows an $E(g; \bs{0}, \bs{I}_{p\times p})$ (spherical) distribution and $\bs{I}_{p\times p}$ is the covariance matrix of $(\bs{x}', y)$.

\vs
\noin
\tb{Remarks 6.1}

\noin
\tb{(I)} Stopping criteria for the algorithm include (i) the total number of the LS estimation decided to perform (ii) the total number of two indices sampled from $\{1,2, \cdots, n\}$ or (iii) the total number of distinct
index sequences $i_1, \cdots, i_K$ in the step (a2) of  (3).
\vs
\tb{(II)} 
There are  $O(n^2)$ two-point pairs, all other operations cost at most $O(np^2+p^3)$, theoretically, overall the worst time complexity is $O(pn^3+n^2p^3)$. However, in the program, $N$ is the minimum of $\{1000, {n \choose \lfloor(n+1)/2\rfloor}, T_{ls}\}$, where $T_{ls}$ is a turning parameter, the total number of the LS estimation decided to perform, which usually set to be $100\sim 500$, so in practice the real time complexity is $O(np^2+p^3)$ (see Section 7).
\vs
\tb{(III)} When $\bs{x}_i=\bs{x}_j$ for some $i\neq j$, one can add a small $\varepsilon$ say, to $\bs{x}_i$, to force them are not identical. So that one can still apply the AA1.
\hfill\pend

\vs
\noindent
\tb{Remarks 6.2}

\noin
\tb{(I)} It is readily seen that the worst case time complexity of algorithm AA2 is $O(N(p^2n+p^3))$ where $p^3$ comes from  finding the inverse of $p$ by $p$ matrix and from $p\times p$ matrix multiply a $p$ vector and  the most costly step is (1) to compute the $I(\bs{\beta}_{new})$ which, however, can  achieve in $O(np^2)$. When $n$ and $p$ are small (say $n \leq 50$, $p \leq 3$), then $N$ might just be ${n \choose p}$, otherwise it will be $300(p-1)$. Here $300$ could be tuned to a larger number - such as $500$ - or even larger.  It is readily seen that the AA2 produces a non-negative and non-increasing sequence: $Q_1> Q_2\cdots > Q_k> \cdots $. So the convergence of AA2 is always achievable. \vs

\tb{(II)} For large $n$, say $n\geq 200$, we suggest that one first partitions the data set into disjoint (say five) subsets, then applies the AA2 to each subset to obtain $\bs{\beta}$ from each subset. Finally, one
carries out step (1) above with respect to the entire data set and selects the $\bs{\beta}$ which produces the smallest objective function value $Q(\bs{\beta})$.
\vs
\tb{(III)} In the algorithm AA2, the sub-sample size $m$ is $p$. Other choices include $\lfloor (n+1)/2 \rfloor$ (corresponding to $\alpha=1$) and  $I(\bs{\beta}_{new})$ (which requires an initial $\bs{\beta}_{new}$).
The latter however is generally not recommended.
\hfill \pend
\end{description}
\end{document}